\documentclass[a4paper,11pt]{article}
\usepackage{jheppub} % for details on the use of the package, please see the JINST-author-manual
\makeatletter
\gdef\@fpheader{Prepared for submission to JHEP\hfill IPPP/26/87}
\usepackage{lineno}
\usepackage{bm}
\usepackage{graphicx}
\usepackage{subcaption}
\title{\boldmath Probing scalar-mediated Lorentz violation with hydrogen masers and spin-precession of molecules}

\author{Djuna Croon}
\author{and Tanmay Kumar Poddar}

\affiliation{Institute for Particle Physics Phenomenology (IPPP),
Department of Physics, Durham University,
Durham DH1 3LE, United Kingdom}

\emailAdd{djuna.l.croon@durham.ac.uk}
\emailAdd{tanmay.k.poddar@durham.ac.uk}

\abstract{We propose using hydrogen maser spectroscopy and molecular spin-precession experiments, traditionally employed in precision frequency measurements and searches for the electron electric dipole moment, as complementary probes of light scalar-mediated, spin-dependent fifth forces. We investigate their sensitivity to Lorentz-violating interactions mediated by scalars with masses ranging from $\mathcal{O}(10^{-4}~\mu\mathrm{eV})$ to $\mathcal{O}(10~\mathrm{keV})$. While hydrogen maser spectroscopy and molecular spin-precession measurements probe different combinations of Lorentz-violating couplings, we find that the strongest projected sensitivities can be achieved using spin-precession of ultracold molecules, benefiting from their extended coherence times. These experimental approaches provide access to previously unexplored regions of parameter space, probing interaction ranges extending from centimeter to sub-angstrom scales. Our proposals highlight the potential of precision atomic and molecular experiments to search for new spin-dependent scalar-mediated interactions and violations of fundamental spacetime symmetries.}

\begin{document}
\maketitle
\flushbottom

\section{Introduction}
Lorentz and CPT (Charge conjugation-Parity-Time reversal) invariance are fundamental symmetries underlying the Standard Model (SM) of particle physics and general relativity (GR). Nevertheless, several approaches to physics beyond the SM, particularly those attempting to incorporate gravity at the quantum level, allow for small violations of these symmetries at experimentally accessible energies \cite{Kostelecky:1988zi,Colladay:1996iz}. Such effects may arise, for example, through spontaneous Lorentz-symmetry breaking in an underlying high energy theory \cite{Kostelecky:1988zi,Kostelecky:1991ak}. A systematic effective field theory (EFT) framework for describing such effects is provided by the Standard Model Extension (SME) \cite{Colladay:1996iz,Colladay:1998fq}, which supplements the SM and GR with all possible coordinate independent Lorentz and CPT-violating operators. In EFT, CPT violation necessarily implies Lorentz violation (LV), whereas LV need not in general imply CPT violation \cite{Greenberg:2002uu}. The SME therefore provides a unified framework for systematically confronting possible Lorentz and CPT-violating effects with precision experiments.

While Lorentz and CPT violating effects involving the fermion and gauge sectors have been extensively investigated \cite{Colladay:1998fq,Kostelecky:2002hh,Heckel:2008hw,Kostelecky:2009zp,Stanwix:2005yv}, comparatively less attention has been devoted to interactions involving scalar fields \cite{Altschul:2012xu}. This is particularly interesting because scalar degrees of freedom arise naturally in the SM and in numerous extensions of it and can mediate new interactions between ordinary matter. Scalar couplings to ordinary matter also arise in models of CP-violating axions, with implications for fifth-force searches and astrophysical field configurations \cite{OHare:2020wah,Anzuini:2023whm,Ramadan:2024vfc}. In the presence of Lorentz and/or CPT violating scalar-fermion couplings, scalar exchange can generate modified Yukawa-type interactions with non-trivial spin and directional dependence. At non-relativistic energies of fermions, such interactions therefore provide a natural connection between Lorentz and CPT violating EFTs and precision searches for new scalar-mediated spin-dependent force.

A particularly interesting possibility arises when LV manifests through the interactions of a new light scalar field with SM fermions, rather than through modifications of their free-particle dispersion relations. The most general renormalizable scalar-fermion interaction, including Lorentz- and CPT-violating operators, can be expressed as \cite{Altschul:2012xu}

\begin{equation}
\mathcal{L}_{\rm{int}}=-\phi\sum_F\overline{\psi}_F[g_F+ig^\prime_F\gamma_5+I_\mu^F\gamma^\mu+J^F_\mu\gamma_5\gamma^\mu+\frac{1}{2}L^F_{\mu\nu}\sigma^{\mu\nu}]\psi_F,
\label{eq:1}
\end{equation}

where $\phi$ denotes the scalar field and $\psi_F$ represents an SM fermion of species $F$. The dimensionless coefficients $g_F$ and $g^\prime_F$ describe the conventional Lorentz-invariant scalar and pseudoscalar interactions, respectively, whereas $I_\mu^F$, $J_\mu^F$, and $L_{\mu\nu}^F$ parametrize Lorentz-violating vector, axial-vector, and tensor couplings. Among these, the operators associated with $I_\mu^F$ and $J_\mu^F$ are also CPT-odd, thereby introducing CPT violation.

The exchange of the scalar mediator between fermions generates modified Yukawa-type interactions, whose structure depends on the Lorentz-invariant and Lorentz-violating couplings at the interaction vertices. In the non-relativistic limit, these interactions give rise to both spin-independent and spin-dependent potentials, exhibiting characteristic spatial, angular, and directional dependence \cite{Altschul:2012xu}. The interaction range is governed by the scalar mass $m_\phi$, with a characteristic Compton wavelength $\lambda_\phi\sim m_\phi^{-1}$. Consequently, precision measurements of spin-dependent energy shifts provide a promising avenue for probing such interactions, particularly those involving mixed scalar and Lorentz-violating couplings at the two vertices of the scalar-exchange diagram.

Macroscopic torsion-balance and magnetometer experiments have been employed to constrain exotic spin-dependent Lorentz-violating interactions mediated by ultralight scalar particles \cite{Carenza:2025jwn}. Their sensitivities are significantly enhanced by the coherent contribution of a large number of particles in macroscopic sources, including the Earth and the Sun, provided that the interaction range is sufficiently long. However, as the mediator mass increases, the Yukawa suppression becomes increasingly important when the scalar Compton wavelength falls below the characteristic source-detector separation, substantially reducing the sensitivity of these experiments. This motivates complementary precision measurements at shorter length scales, including atomic spectroscopy and spin-precession experiments with atoms and molecules, which can probe scalar-mediated Lorentz-violating interactions over different mediator mass ranges.

In the non-relativistic limit of the fermions, a particularly simple spin-dependent interaction is controlled by $\tilde{J}^F_j=J_j^F+\epsilon_{jkl}L^F_{kl}$. Exchange of a scalar boson between an unpolarized fermion (B) with scalar-like interaction and a spin-polarized fermion (A) with Lorentz-violating spin-dependent interaction generates the potential $V_{\tilde{J}}(r)=-g_B(\bm{\sigma}_A\cdot \tilde{\bm{J}}^A)(e^{-m_\phi r}/4\pi r)$.

Constraints on the coupling $g\tilde{J}$ have previously been derived in \cite{Carenza:2025jwn} for scalar masses $m_\phi \ll 10^{-10}~\mathrm{eV}$, using bounds on the Standard Model Extension (SME) parameters obtained from torsion-balance and magnetometer experiments. A novel strategy for probing such mixed couplings using Josephson junctions has been proposed for scalar masses $10^{-6}~\mathrm{eV}\lesssim m_\phi \lesssim 5~\mathrm{eV}$ \cite{Croon:2026ake}. In addition, spectroscopic measurements in atoms and highly charged ions have been employed to constrain alternative classes of mixed couplings \cite{Jaeckel:2026aeh}. Among these techniques, hydrogen maser spectroscopy \cite{Kleppner:1962zz,PhysRev.138.A972} is particularly attractive owing to its exceptional frequency stability and sensitivity to tiny shifts in hyperfine and Zeeman energy levels.

Originally developed as a highly stable atomic frequency standard based on the ground state hyperfine transition of hydrogen, hydrogen maser spectroscopy has subsequently found widespread applications in precision tests of fundamental physics such as tests of GR \cite{Vessot:1980zz}, the equivalence principle \cite{Ashby:2007zz}, variations of fundamental constants \cite{Prestage:1995zz}, ultralight dark matter (DM) \cite{Kennedy:2020bac}, quantum electrodynamics (QED) \cite{Eides:2000xc}, proton structure \cite{Dupays:2003zz}and Lorentz- and CPT-violating interactions \cite{Phillips:2000dr,Humphrey:2001wm}. In particular, precise measurements of the Zeeman transition frequencies of the hydrogen levels offer a sensitive probe of spin-dependent Lorentz violation, which can manifest as characteristic sidereal modulations due to the Earth's rotation relative to a preferred direction in space.

In this work, we investigate scalar-mediated Lorentz-violating interactions between the electron and proton within a hydrogen atom and explore their signatures through sidereal variations of the Zeeman transition frequency. Unlike macroscopic long-range force experiments, which primarily probe ultralight mediators, atomic spectroscopy provides access to a complementary scalar mass range \cite{Carenza:2025jwn}. The characteristic electron-proton separation in hydrogen is determined by the Bohr radius, $a_H=(\alpha\mu)^{-1}$, where $\alpha$ is the fine-structure constant and $\mu=m_em_p/(m_e+m_p)$ is the electron-proton reduced mass, with $m_e$ and $m_p$ denoting the electron and proton masses, respectively. Numerically, $a_H^{-1}\simeq 3.727~\mathrm{keV}$, indicating that hydrogen maser spectroscopy can probe scalar-mediated interactions at characteristic mediator masses of order a few keV, substantially extending the mass reach beyond that of conventional macroscopic fifth-force searches.

We also extend the above strategy to probe Lorentz-violating interactions mediated by a light scalar field, sourced by an external source. An external unpolarized macroscopic source generates the scalar field through its ordinary scalar coupling to matter, while the scalar couples to the electron and/or proton spin of hydrogen atom through Lorentz-violating operators. The resulting interaction produces a source-dependent perturbation of the hydrogen Zeeman levels. By periodically varying the distance between the source and the maser, the interaction strength is modulated at a source modulation frequency along with two characteristic sidebands due to the sidereal motion of the Earth. This source-correlated spectral sideband structure provides an additional discriminator against conventional static backgrounds, axion-mediated monopole-dipole force \cite{Crescini:2016lwj} and motivates the modulated source hydrogen maser search. Such a setup is sensitive to scalar masses set by the inverse characteristic separation between the external source and the detector. For a typical source to maser-bulb distance of $\mathcal{O}(10~\mathrm{cm})$, the corresponding mass scale is $\mathcal{O}(\mu\mathrm{eV})$, allowing the experiment to probe scalar mediators in the $\mu \rm{eV}$ mass range.

Complementary to hydrogen maser Zeeman spectroscopy, precision spin-precession measurements using atoms and molecules offer a powerful approach to probing fundamental symmetries and searching for weak spin-dependent interactions, including Lorentz- and CPT-violating effects \cite{Vasilakis:2008yn,Agrawal:2023lmw}. In these experiments, coherent superpositions of spin states accumulate a relative phase during free precession, which can be measured with high precision using Ramsey interferometry \cite{Ramsey:1950tr}. Molecular systems are particularly promising owing to their efficient state preparation and readout, potentially long coherence times, and the availability of internal co-magnetometer states in suitable molecules, which help suppress magnetic field induced systematic effects. These capabilities have enabled sensitive searches for electric dipole moments and motivated new strategies for probing exotic spin-dependent forces and ultralight DM \cite{ACME:2018yjb,Anderegg:2023dgq,Graham:2013gfa}.

Recently, atomic and molecular spin-precession experiments have been proposed to search for axion mediated monopole-dipole interactions, with ACME III like thorium monoxide (ThO) molecular beam experiments providing a representative benchmark \cite{Agrawal:2023lmw}. A projected precession frequency sensitivity of $\delta\omega_{\rm exp}\sim15~\mu\mathrm{rad/s}$ has been considered for a nearby dense lead or tungsten source with characteristic dimensions and separation of $\mathcal{O}(10~\mathrm{cm})$. Motivated by this approach, we investigate scalar mediated Lorentz violating interactions in which a light scalar couples Lorentz invariantly to ordinary matter in an unpolarized macroscopic source and through a Lorentz violating spin-dependent coupling to electrons in the atomic or molecular sensor. The resulting scalar field induces an additional spin-dependent energy splitting, leading to an anomalous precession phase, $\Phi_{\rm LV}\simeq\delta\omega_{\rm LV}\tau$, accumulated over the interrogation time $\tau$ when the interaction is approximately constant during each measurement. Periodic modulation of the source position can further generate a characteristic time-dependent signal, providing a means of distinguishing the new interaction from conventional magnetic and static gravitational backgrounds.

Additionally, advances in the cooling, trapping, and coherent control of ultracold molecules have opened new opportunities for precision searches for physics beyond the SM \cite{Anderegg:2023dgq}. Unlike molecular beam experiments, trapped ultracold molecules can remain localized near an external source for extended periods, overcoming transit time limitations and enabling longer coherent interrogation times \cite{Hallas:2022rdw,Vilas:2023nhd,Ruttley:2024dms,Robichaud2026,Park:2016aqi}. This enhances the accumulated spin precession phase induced by weak spin-dependent interactions and can substantially improve sensitivity to anomalous precession frequencies. Furthermore, improved control over molecular quantum states and reduced spatial averaging of the source-induced interaction make ultracold molecules particularly promising for probing scalar-mediated Lorentz-violating couplings.

The remainder of this paper is organized as follows. In Section~\eqref{newsec2}, we briefly review hydrogen maser spectroscopy and introduce the Sun-centered reference frame used to characterize sidereal modulations induced by LV. In Section~\eqref{sec2}, we investigate scalar-mediated LV interactions between the electron and proton within a hydrogen atom and derive constraints on the corresponding mixed couplings from sidereal variations of the Zeeman transition frequency. In Section~\eqref{sec3}, we extend the analysis to an external macroscopic scalar source and determine the sensitivity to LV couplings through the resulting Zeeman-frequency modulations. In Section~\eqref{molecule-section}, we explore complementary constraints from molecular spin-precession measurements and potential sensitivity improvements achievable with ultracold molecules. Finally, Section~\eqref{sec4} summarizes our findings and discusses their implications. Throughout this work, we adopt natural units, $\hbar=c=1$, unless otherwise stated.

\section{Hydrogen maser spectroscopy and the choice of reference frame}\label{newsec2}

\begin{figure}
    \centering
    \includegraphics[width=10cm]{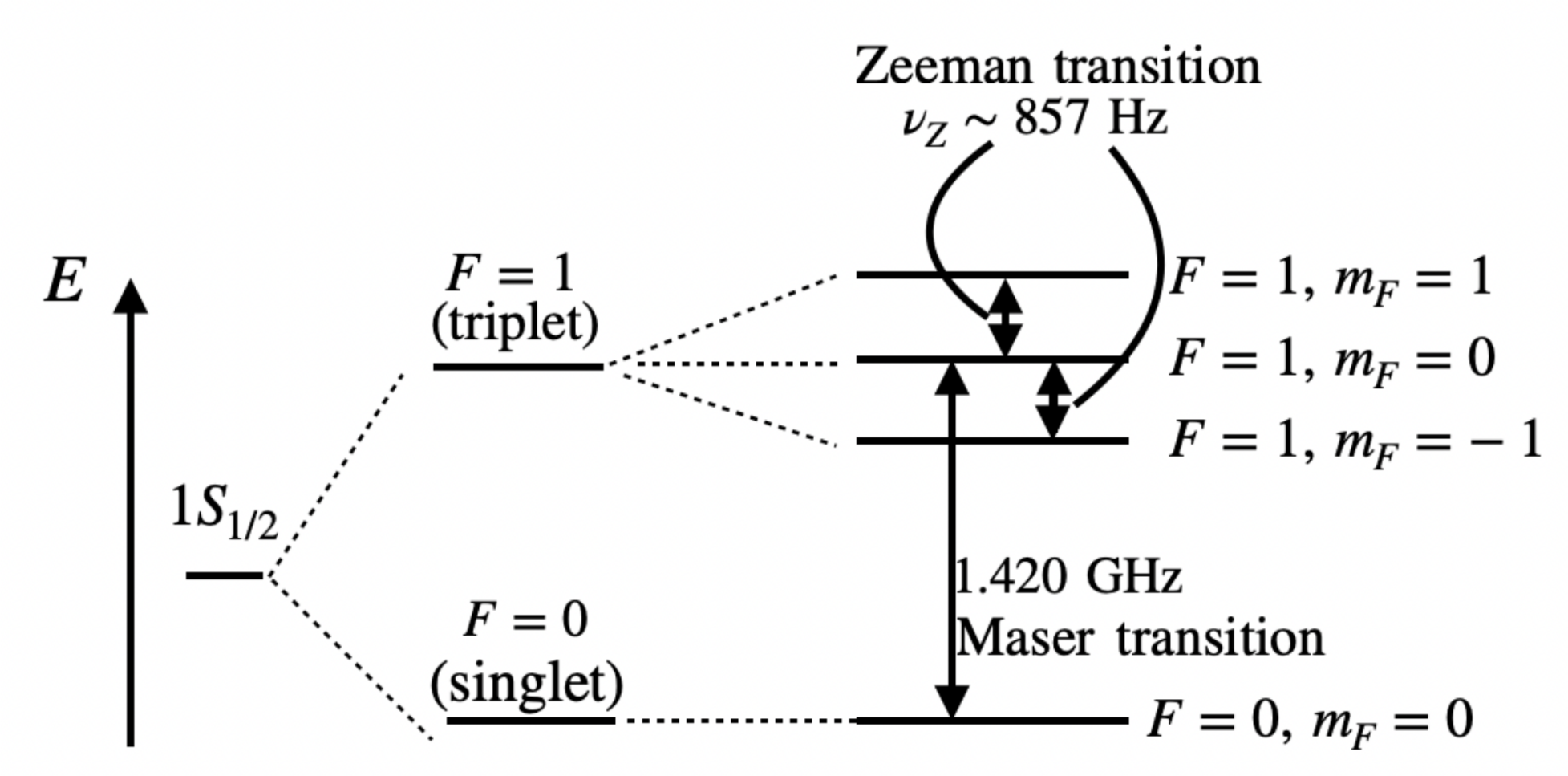}
    \caption{Ground state atomic hydrogen (1S) hyperfine and Zeeman levels.}
    \label{fig1}
\end{figure}

Hydrogen maser spectroscopy provides a sensitive probe of spin-dependent Lorentz violating interactions owing to the exceptionally narrow and well-controlled hyperfine structure of ground state atomic hydrogen. The $1S_{1/2}$ ground state is split by the hyperfine interaction into an $F=1$ triplet and $F=0$ singlet as shown in FIG.~\eqref{fig1}. In the presence of the weak static $C-$ (magnetic) field used in maser operation, the triplet is further resolved into the $m_F=+1,0,-1$ Zeeman sublevels. The maser oscillates on the $|F=1,m_F=0\rangle\leftrightarrow |F=0,m_F=0\rangle$ transition at $\nu_{\mathrm{hfs}}\simeq 1.4204~\mathrm{GHz}$, which is insensitive to the magnetic field at first order and sensitive to the second order as $\nu_{24}=\nu_{\mathrm{hfs}}+2750B^2$, where $\nu$ is taken in $\rm{Hz}$ units and $B$ is in Gauss.
In contrast, the adjacent $F=1, \Delta m_F=\pm 1$ transitions are linearly sensitive to the applied magnetic field and occur at a characteristic Zeeman frequency of order $857~\mathrm{Hz}$.

For weak magnetic fields, the two adjacent Zeeman frequencies are $\nu_{12}=1.4\times 10^6 B-1375B^2$ and $\nu_{23}=1.4\times 10^6 B+1375B^2$. The mean $\nu_Z=(\nu_{12}+\nu_{23})/2\simeq 1.4\times 10^6 B$, provides the experimentally measured Zeeman resonance. For the operating field $B=B_C\simeq 0.61~\mathrm{mG}$, we obtain $\nu_Z\simeq 857~\mathrm{Hz}$, while the separation of the two transitions $\nu_{12}-\nu_{23}\simeq 3~\mathrm{mHz}$ is much smaller than the Zeeman linewidth of $\sim 1~\mathrm{Hz}$. The two transitions therefore form an effectively unresolved Zeeman resonance that can be be accurately determined by applying a weak transverse rf magnetic field and employing double resonance spectroscopyv \cite{Humphrey:2001wm}. The rf field drives transitions between the Zeeman sublevels, thereby perturbing the microwave maser oscillation frequency. By sweeping the rf frequency across the resonance and monitoring the resulting shift in the maser frequency, the Zeeman resonance frequency can be accurately extracted.

The total angular momentum of the hydrogen atom is given as $\bm{F}=\bm{I}+\bm{J}$, where $\bm{I}$ is nuclear spin angular momentum and $\bm{J}=\bm{L}+\bm{S}$ is the total electronic angular momentum, where $\bm{L}$ is the electron orbital angular momentum and $\bm{S}$ is the electron spin. For the hydrogen ground state $1S$, which contains one electron, $L=0,~S=1/2$ and for the ordinary hydrogen atom, the nucleus contains one proton, which corresponds to $I=1/2$. Therefore, the possible hyperfine levels are $F=|I-J|,\cdots I+J\equiv 0,1$. Once, $F$ is fixed, quantum mechanics allows the projection $F_z=m_F$, with $(m_F=-F,-F+1\cdots F)$ $2F+1$ magnetic sublevels. The $F=1$ state is triplet manifold containing three $(m_F=-1,0,1)$ magnetic sublevels. Similarly, for $F=0$ singlet state, $m_F=0$ is possible. Therefore, the magnetic sublevels are denoted by $|F,m_F\rangle$.

The hydrogen ground state hyperfine manifold $F=1,0$ contains four ground state hyperfine sublevels $|F,m_F\rangle$ given as
\begin{align}
|1,1\rangle &=|\uparrow_e\uparrow_p\rangle,\nonumber\\
|1,0\rangle&=\frac{1}{\sqrt{2}}(|\uparrow_e\downarrow_p\rangle+|\downarrow_e\uparrow_p\rangle),\nonumber\\
|1,-1\rangle&=|\downarrow_e \downarrow_p\rangle,\nonumber\\
|0,0\rangle&=\frac{1}{\sqrt{2}}(|\uparrow_e\downarrow_p\rangle-|\downarrow_e\uparrow_p\rangle).
\label{eq:8}
\end{align}
The hyperfine splitting is produced by the internal magnetic interaction between the electron and proton magnetic moments. The energy difference between these two hyperfine levels is $\Delta E_{\rm{hfs}}=2\pi \nu_{\rm{hfs}}\simeq 1.42~\mathrm{GHz}$, which corresponds to the $21~\mathrm{cm}$ hydrogen line. Although the spontaneous magnetic dipole transition between these levels has an exceptionally long lifetime of approximately $10^7~\mathrm{years}$, a resonant microwave field can induce transitions through stimulated emission. In a hydrogen maser, a population inversion between the hyperfine levels enables stimulated emission to sustain a highly stable and coherent microwave signal at approximately $1.42~\mathrm{GHz}$.

At zero magnetic field, the magnetic sublevels for $F=1$ state are $|1,1\rangle, ~ |1,0\rangle,~|1,-1\rangle$, which have same energy and hence degenerate. To break this degeneracy, a weak static magnetic field, commonly known as the $C-$ field can be applied, which defines the quantization axis, say $\bm{B}_C=B_C\hat{z}$, where $\hat{z}$ is the laboratory $z-$ axis, fixed to the apparatus. Therefore, the corresponding magnetic Zeeman interaction term $H_{\rm{Zeeman}}=-\bm{\mu}\cdot \mathbf{B_C}=-\mu_z B_C$. The first order perturbative energy shift becomes $\Delta E_{\rm{Zeeman}}=\langle F,m_F|H_{\mathrm{Zeeman}}|F,m_F\rangle=g_F\mu_B m_F B_C$, where $g_F$ is the hyperfine Land\'e $g$-factor and $\mu_B=e/2m_e$ is the Bohr magneton with $e$, the electronic charge and $m_e$, the mass of the electron and hence, $m_F=-1,0,1$ acquire different energies. The expression for the energy shift is valid in the weak field regime, where the Zeeman interaction is much smaller than the $1.42~\mathrm{GHz}$ hyperfine splitting and $F, m_F$ remain good quantum numbers.

The transitions relevant to hydrogen maser spectroscopy obey the magnetic dipole (M1) selection rules $\Delta F=0,\pm1$, with $F=0\leftrightarrow F=0$ forbidden, and $\Delta m_F=0,\pm1$, depending on the polarization of the applied oscillating magnetic field. The conventional hyperfine transition $|1,0\rangle\leftrightarrow|0,0\rangle$, corresponding to a frequency of approximately $1.42~\mathrm{GHz}$, satisfies $\Delta F=\pm1$ and $\Delta m_F=0$. In contrast, the Zeeman transitions $|1,1\rangle\leftrightarrow|1,0\rangle$ and $|1,0\rangle\leftrightarrow|1,-1\rangle$ satisfy $\Delta F=0$ and $\Delta m_F=\pm1$ and are driven by a weak transverse rf magnetic field. A weak static magnetic field, commonly referred to as the $C-$field, with strength $B_C\simeq0.61~\mathrm{mG}$, defines the quantization axis and lifts the degeneracy of the $F=1$ magnetic sublevels, resulting in a Zeeman transition frequency of approximately $\nu_{\mathrm{Zeeman}}\simeq857~\mathrm{Hz}$ \cite{Phillips:2000dr}.

\begin{figure}
    \centering
    \includegraphics[width=12cm]{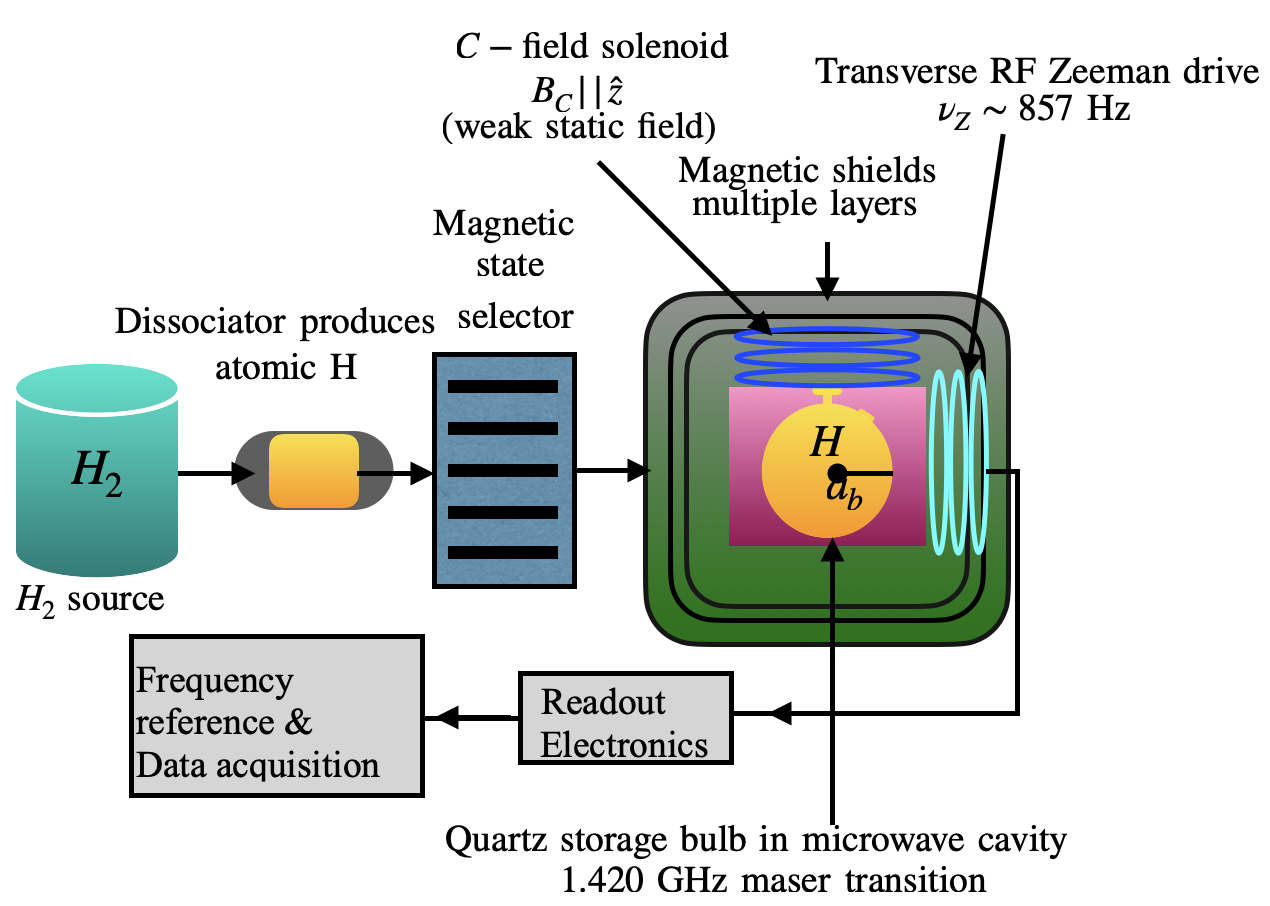}
    \caption{Hydrogen maser apparatus for probing scalar-mediated Lorentz-violating interactions between the electron and proton within an atomic hydrogen system.}
    \label{fig0}
\end{figure}

Figure.~\eqref{fig0} shows the schematic experimental configuration for hydrogen maser. Molecular hydrogen from the ${\rm H}_2$ source is first dissociated to produce a beam of atomic hydrogen. A magnetic state selector, which does the population inversion, preferentially focuses the appropriate low field seeking hyperfine states into the quartz storage bulb, which is located inside a microwave cavity resonant with the $1.420~\mathrm{GHz}$ ground state hyperfine transition. A weak static magnetic field $B_C$ generated by the $C-$ field solenoid, defines the laboratory quantization axis and lifts the degeneracy of the $F=1$ Zeeman sublevels. A transverse rf magnetic field drives the $F=1,~\Delta m_F=\pm 1$ Zeeman transitions, whose characteristic resonance frequency is $\nu_Z\simeq 857~\mathrm{Hz}$ for the operating magnetic field considered here. The cavity and field coils are enclosed within multiple layers of high permeability magnetic shielding in order to suppress environmental magnetic field fluctuations. The Zeeman resonance is determined through the double resonance technique by monitoring its effect on the maser signal, which is subsequently processed by the readout electronics and frequency-reference/data acquisition system.

\begin{figure}
    \centering
    \includegraphics[width=12cm]{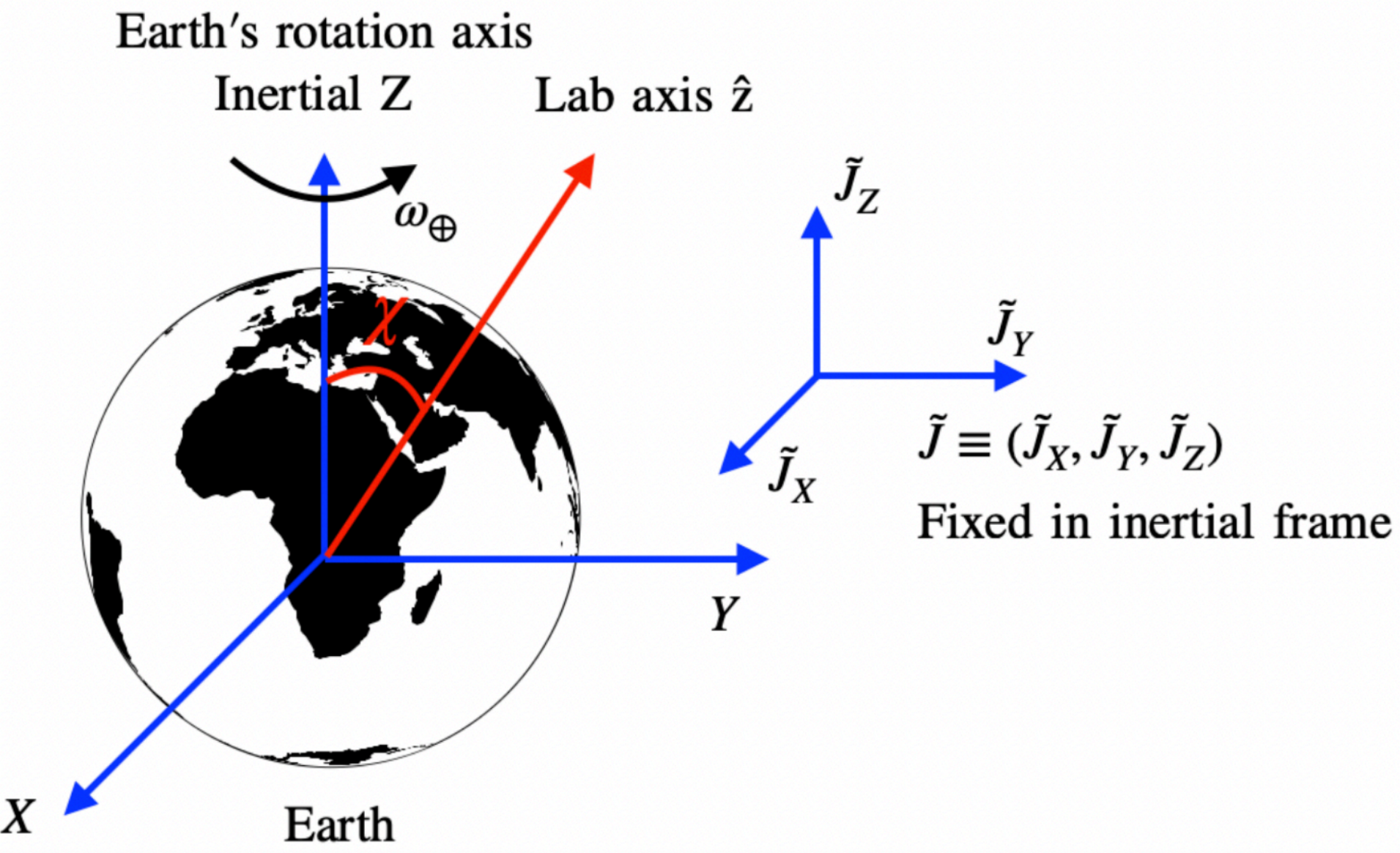}
    \caption{Schematic representation of the orientation of the laboratory quantization axis relative to the Sun-centered celestial-equatorial frame. For clarity, the Sun-centered celestial equatorial frame axes are translated to the Earth's center while retaining their orientation; their actual origin is conventionally taken at the Sun.}
    \label{fig2}
\end{figure}

To investigate the sidereal modulation induced by Lorentz-violating interactions, we adopt the conventional Sun-centered celestial equatorial frame (SCCEF), with spatial coordinates $(X,Y,Z)$. The origin of this approximately inertial frame is located at the center of the Sun, with the $Z$-axis aligned with the Earth's rotation axis and directed towards the celestial north pole $(\hat{Z}\parallel\boldsymbol{\omega}_\oplus)$. The $X$-axis points towards the vernal equinox, while the $Y$-axis completes the right-handed orthonormal coordinate system, $\hat{Y}=\hat{Z}\times\hat{X}$.\footnote{Although the SCCEF is not strictly inertial due to the acceleration of the Sun within the Solar System and the Galaxy, these effects are negligible on the timescales relevant to terrestrial sidereal-modulation experiments. The frame can therefore be treated as quasi-inertial for the present analysis.} In the laboratory frame, the local $z$-axis is chosen along the vertical direction, and $\chi$ denotes the colatitude of the experimental location, defined as the angle between the local vertical and the Earth's rotation axis.

Figure~\eqref{fig2} illustrates the orientation of the laboratory quantization axis with respect to the Sun-centered celestial-equatorial frame. The Lorentz-violating background, $\widetilde{\mathbf J}=(\widetilde J_X,\widetilde J_Y,\widetilde J_Z)$, is taken to be fixed in this approximately inertial frame, whose $Z$-axis is aligned with the Earth's rotation axis. The laboratory axis $\hat z$, defined by the static $C$-field of the hydrogen maser, makes an angle $\chi$ with the $Z$-axis and rotates with the Earth at the sidereal frequency $\omega_\oplus$. Consequently, the projection of the fixed LV background onto the laboratory quantization axis acquires a characteristic sidereal time dependence, providing the basis for the modulation search.

The sensitivities to Lorentz-violating couplings are initially expressed in the laboratory frame, where the corresponding coefficients exhibit an apparent time dependence due to the Earth's rotation. To characterize this sidereal modulation, we transform the laboratory frame coefficients into the approximately inertial Sun-centered celestial equatorial frame, in which the underlying Lorentz-violating background coefficients are conventionally taken to be constant. 

As discussed below, we express the sensitivities to Lorentz-violating couplings derived from hydrogen maser spectroscopy and molecular spin-precession measurements in the Sun-centered reference frame.

\section{Hydrogen-maser probe of scalar-mediated Lorentz violation}\label{sec2}

We consider scalar-mediated Lorentz-violating interactions between the electron ($e$) and proton ($p$) in a hydrogen atom, allowing for both Lorentz-invariant and Lorentz-violating couplings at the respective interaction vertices. The corresponding interaction Lagrangian is given as
\begin{equation}
\mathcal{L} \supset -\phi\sum_{F=e,p}\overline{\psi}_F[g_F+J^F_\mu\gamma_5\gamma^\mu+\frac{1}{2}L_{\mu\nu}^F\sigma^{\mu\nu}] \psi_F, 
\label{eq:2}
\end{equation}
where $g_F$ denotes the Lorentz-invariant scalar coupling, while $J_\mu^F$ and $L_{\mu\nu}^F$ characterize the Lorentz-violating axial-vector and tensor couplings, respectively. The axial-vector coefficients $J_\mu^F$ additionally violate CPT symmetry. In the non-relativistic limit, the exchange of the scalar mediator between the electron and proton generates spin-dependent interactions arising from mixed scalar and Lorentz-violating couplings at the two vertices. Retaining terms to first order in the Lorentz-violating coefficients, the resulting electron-proton potential can be expressed as
\begin{equation}
V_{\mathrm{LV}}(r)=\Big[g_p(\bm{\sigma}_e\cdot \tilde{\bm{J}}^e)+g_e (\bm{\sigma}_p\cdot \tilde{\bm{J}}^p)\Big] f(r),~~~f(r)=-\frac{e^{-m_\phi r}}{4\pi r},
\label{eq:5}
\end{equation}
where $\tilde{J}^{(e,p)}_i=J_i^{(e,p)}+\epsilon_{ijk}L^{(e,p)}_{jk}$, $\sigma_{e,p}$ are the Pauli spin operators, and the sensitivity on the mixed coupling coefficients $(g_{(p,e)}\tilde{J}_i^{(e,p)})$ can be obtained from hydrogen maser spectroscopy in the Sun-centered approximately inertial frame, where $i=X,Y,Z$ are the components of $\tilde{J}_i^{(e,p)}$ in that frame. 

To study the scalar-mediated Lorentz-violating interaction between the electron and proton of an hydrogen atom, we have to calculate the radial matrix element for the hydrogen $1S$ ground state as the hydrogen electron is described by a wavefunction and the separation $r$ between electron and proton is not a fixed classical distance. We write the normalized hydrogen $1S$ wave function as
\begin{equation}
\psi_{1S}=\frac{1}{\sqrt{\pi a^3_H}}e^{-r/a_H}. 
\label{eq:6}
\end{equation}
Since, the $1S$ state is spherically symmetric, we can write the expectation value of the Yukawa function as
\begin{equation}
\mathcal{F}(m_\phi)=\langle f(r)\rangle=-\frac{1}{4\pi}\int d^3 r|\psi_{1S}(r)|^2\frac{e^{-m_\phi r}}{r}=-\frac{1}{4\pi a_H}\frac{1}{\Big(1+\frac{m_\phi a_H}{2}\Big)^2},
\label{eq:7}
\end{equation}
where the characteristic mass scale appears in Eq.~\eqref{eq:7} is $2/a_H\sim 7.45~\mathrm{keV}$. The energy shift of the hydrogen $1S$ state $\delta E=\langle\psi_{1S}|V_{\mathrm{LV}}|\psi_{1S}\rangle$ can be calculated using Eqs.~\eqref{eq:5} and~\eqref{eq:7}, which we obtain by averaging over the full probability distribution of electron-proton separation. 

In the limit $m_\phi a_H\ll 1$, the range of the scalar-mediated force is larger than the atom, and hence $\mathcal{F}\to -1/4\pi a_H$. In the opposite limit $m_\phi a_H\gg 1$, the interaction range is much shorter than the atomic size, and hence $\mathcal{F}\to -1/\pi a^3_H m^2_\phi$. Therefore, the atomic matrix element is essentially independent of $m_\phi$ for $m_\phi\ll \mathrm{keV}$ and decreases only algebraically for heavier mediators.

For hydrogen maser LV measurements the $\Delta m_F=\pm 1$ transitions are important. This is because the Lorentz-violating spin interaction shifts the $m_F=+1$ and $m_F=-1$ in opposite directions. In contrast, the standard $1.42~\mathrm{GHz}$ hyperfine transition ($\Delta m_F=0$) is not affected by spin-dependent Lorentz-violating interactions at the leading order, as the $m_F=0$ states have zero net spin projection. Thus, for our case of scalar-mediated spin-dependent LV, the Zeeman transition is an unsuppressed first order probe. We do note that other LV operators, momentum-dependent terms, tensor couplings, or second order mixing may shift the $1.42~\mathrm{GHz}$ line. 

Thus, with the choice of the laboratory quantization axis along the $z$ direction, the spin-projections are obtained as
\begin{align}
\langle 1,1|\sigma^e_z|1,1\rangle&=\langle 1,1|\sigma_z^p|1,1\rangle =+1,\nonumber\\
\langle 1,-1|\sigma^e_z|1,-1\rangle&=\langle 1,-1|\sigma_z^p|1,-1\rangle =-1,\nonumber\\
\langle 1,0|\sigma^e_z|1,0\rangle&=\langle 1,0|\sigma_z^p|1,0\rangle =0.
\label{eq:9}
\end{align}
Therefore, the first order Lorentz-violating energy shifts are
\begin{align}
\delta E_{+1}&=\mathcal{F}(m_\phi)(g_p\tilde{J}^e_z+g_e\tilde{J}^p_z),\nonumber\\
\delta E_0&=0,\nonumber\\
\delta E_{-1}&=-\mathcal{F}(m_\phi)(g_p\tilde{J}^e_z+g_e\tilde{J}^p_z).
\label{eq:10}
\end{align}

For states quantized along the laboratory-$z$ axis, the transverse Pauli operators $\sigma_x$ and $\sigma_y$ have vanishing diagonal matrix elements and therefore do not contribute to the first order energy shift. They instead induce off-diagonal mixing between magnetic sublevels. Nevertheless, the Sun-centered coefficients $\tilde{J}_X$ and $\tilde{J}_Y$ contribute at first order because Earth's rotation projects these inertial frame components onto the laboratory-$z$ axis, making $\tilde{J}_z(t)$ time-dependent.
\subsection{Constraints on $g_{p,e}\tilde{J}^{e,p}_{X,Y}$}

\begin{figure}
    \centering
    \includegraphics[width=1\linewidth]{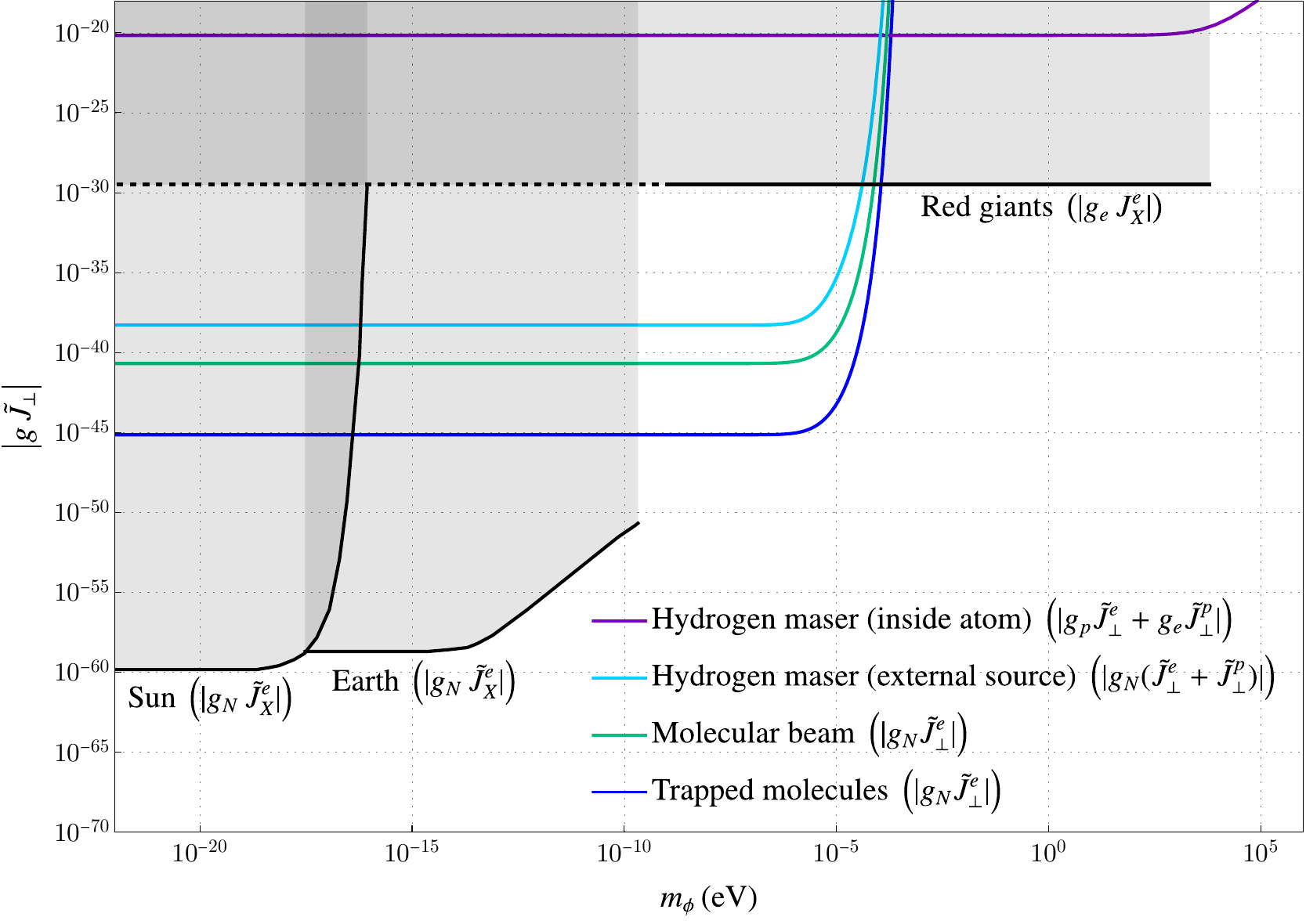}
    \caption{Summary of the projected sensitivities to Lorentz-violating coupling along the direction perpendicular to the Earth's rotation axis. The gray-shaded
    regions represent existing (although on different combination of couplings, discussed in the text) constraints \cite{Carenza:2025jwn}, while the unshaded regions with the purple (hydrogen maser without external source, cf. Sec.~\ref{sec2}), sky-blue (hydrogen maser with external source, cf. Sec.~\ref{sec3}), green (molecular beam, cf. Sec.~\ref{molecular-beam}) and blue (trapped molecules, cf. Sec.~\ref{ultracold-section}) curves indicate parameter space that   can be explored by the proposed experiments discussed in this work.}
    \label{couplingperp}
\end{figure}

We write the two adjacent $F=1$ Zeeman frequencies as
\begin{equation}
\nu_{+1\to 0}=\frac{E_{+1}-E_{0}}{2\pi},~~\nu_{0\to -1}=\frac{E_0-E_{-1}}{2\pi},
\label{eq:11}
\end{equation}
and the hydrogen maser observable is approximately their average, given as
\begin{equation}
\nu_{\mathrm{Zeeman}}=\frac{1}{2}(\nu_{+1\to 0}+\nu_{0\to -1})=\frac{E_{+1}-E_{-1}}{4\pi}.
\label{eq:12}    
\end{equation}
Since, the energy of the $m_F=0$ state cancels in determining $\nu_{\mathrm{Zeeman}}$, any small magnetic field induced mixing between the $F=1,~m_F=0$ and $F=0,~m_F=0$ levels does not affect the leading result. 

Using Eqs.~\eqref{eq:10} and \eqref{eq:12}, we obtain
\begin{equation}
\delta \nu_{\mathrm{Zeeman}}=\frac{\mathcal{F}(m_\phi)(g_p\tilde{J}^e_z+g_e\tilde{J}^p_z)}{2\pi}.  
\label{eq:13}  
\end{equation}
Experimentally one tracks $\nu^{\rm{meas}}_\mathrm{Zeeman}(t)$ over time and look for any sidereal modulation caused by LV. The transformation of the Lorentz-violating vector from the Sun-centered frame to the laboratory frame make the observable time-dependent i.e.,
\begin{equation}
\nu^{\mathrm{meas}}_{\mathrm{Zeeman}}(t)\simeq 857~\mathrm{Hz} +\delta\nu_\mathrm{Zeeman}(t)+\mathrm{ordinary~drifts},
\label{eq:14}  
\end{equation}
where 
\begin{equation}
2\pi\delta\nu_{\mathrm{Zeeman}}(t)=-\frac{1}{4\pi a_H} \frac{(g_p\tilde{J}^e_z(t)+g_e\tilde{J}^p_z(t))}{\Big(1+\frac{m_\phi a_H}{2}\Big)^2},  
\label{eq:15}  
\end{equation}
using Eq.~\eqref{eq:7}.

The projection of the Lorentz-violating coefficient onto the laboratory $z$-axis is therefore obtained by transforming from the Sun-centered frame, yielding
\begin{equation}
\tilde{J}_z(t)=\sin\chi\Big[\tilde{J}_X\cos(\omega_\oplus t)+\tilde{J}_Y\sin(\omega_\oplus t)\Big]+\tilde{J}_Z\cos\chi,  
\label{eq:16}  
\end{equation}
where $\omega_\oplus=2\pi/\textrm{sidereal day}$ and $1~\textrm{sidereal day}=23~\textrm{h}~56~\textrm{m}$. Defining, 
\begin{equation}
\mathcal{C}_z(t)=(g_p\tilde{J}^e_z(t)+g_e\tilde{J}^p_z(t)) 
\label{dcnew1}
\end{equation}
and comparing Eqs.~\eqref{eq:15} and \eqref{eq:16}, we obtain 
\begin{equation}
2\pi \delta\nu_{\mathrm{Zeeman}}(t)=-\frac{1}{4\pi a_H}\frac{1}{\Big(1+\frac{m_\phi a_H}{2}\Big)^2}\Big[\sin\chi \mathcal{C}_X\cos(\omega_\oplus t)+\sin\chi\mathcal{C}_Y\sin(\omega_\oplus t)+\cos\chi\mathcal{C}_Z\Big],  
\label{eq:17} 
\end{equation}
where $\mathcal{C}_{X,Y,Z}$ are the equivalent components of $\mathcal{C}_z(t)$ in the Sun-centered frame, given as
\begin{equation}
\mathcal{C}_i=g_p \tilde{J}_i^e +g_e \tilde{J}_i^p,~~~i=X,Y,Z.   
\end{equation}
Therefore, the cosine and sine sidereal amplitudes are
\begin{equation}
\mathcal{A}^E_X=\frac{\sin\chi}{4\pi a_H} \frac{\mathcal{C}_X}{\Big(1+\frac{m_\phi a_H}{2}\Big)^2},~~~\mathcal{A}^E_Y=\frac{\sin\chi}{4\pi a_H}   \frac{\mathcal{C}_Y}{\Big(1+\frac{m_\phi a_H}{2}\Big)^2},
\label{eq:18}
\end{equation}
where $\mathcal{A}^E=2\pi \mathcal{A}^\nu$ denotes the corresponding energy amplitudes.

The hydrogen maser search has constrained a sidereal variation of the $F=1$, $\Delta m_F=\pm 1$ Zeeman frequency to $\mathcal{A}^\mathrm{exp}_\nu\simeq 0.37~\mathrm{mHz}$ at the one standard deviation level, which corresponds to $\mathcal{A}^\mathrm{exp}_E=2\pi \mathcal{A}^\mathrm{exp}_\nu=1.53\times 10^{-27}~\mathrm{GeV}$ \cite{Phillips:2000dr,Humphrey:2001wm}. The experiment had colatitude $\chi=48^\circ$ corresponds to $\sin\chi=0.743$. Comparing Eq.~\eqref{eq:18} with $\mathcal{A}^\mathrm{exp}_E$, we obtain
\begin{equation}
\frac{\sin\chi}{4\pi a_H} \frac{\sqrt{\mathcal{C}^2_X+\mathcal{C}^2_Y}}{\Big(1+\frac{m_\phi a_H}{2}\Big)^2} \lesssim\mathcal{A}^\mathrm{exp}_E. 
\label{eq:19}
\end{equation}
Using the values of the Bohr radius, colatitude of the experiment and $\mathcal{A}^\mathrm{exp}_E$, we obtain the $1\sigma$ bound on the Lorentz-violating coefficient as
\begin{equation}
\sqrt{\mathcal{C}^2_X+\mathcal{C}^2_Y}= |\mathcal{C}_\perp|=g_p \tilde{J}_\perp^e +g_e \tilde{J}_\perp^p\lesssim 6.9\times 10^{-21} \Big(1+\frac{m_\phi}{7.45~\mathrm{keV}}\Big)^2. 
\label{eq:20}
\end{equation}

If one coupling is Lorentz-violating at a time, then
\begin{equation}
|g_p\tilde{J}^{e}_{\perp}|, |g_e\tilde{J}^{p}_{\perp}|\lesssim  6.9\times 10^{-21} \Big(1+\frac{m_\phi}{7.45~\mathrm{keV}}\Big)^2.
\label{eq:21}
\end{equation}
The constraint on $\mathcal C_\perp$ is independent of $\mathcal C_Z$, since the latter contributes only a time-independent shift to the Zeeman transition frequency (c.f.Eq.~\eqref{eq:17}) and therefore does not affect the sidereal modulation amplitude.
Therefore, the scalar-mediated Lorentz-violating coupling is constrained to be approximately $6.9\times10^{-21}$ in the light-mediator regime $m_\phi\ll 7.45~\mathrm{keV}$. For scalar masses approaching and exceeding $7.45~\mathrm{keV}$, the finite-range suppression of the interaction becomes important, and the resulting constraint progressively weakens. In FIG.~\eqref{couplingperp}, we present the projected sensitivity to the Lorentz-violating coupling combination $|g_p\tilde{J}^e_\perp+g_e\tilde{J}^p_\perp|$, represented by the solid purple curve. The gray-shaded regions indicate the existing constraints \cite{Carenza:2025jwn} obtained by considering red giants, the Sun, and the Earth as sources of the scalar field. The constraints derived from the Sun and the Earth correspond to the Lorentz-violating coupling combination $|g_N\tilde{J}_X^e|$, whereas the red giant bounds are obtained by multiplying the individual constraints on $g_e$ and $|J_X^e|$. The dashed red giant curve represents the low mass regime, where the long range force mediated by the boson can become comparable to the gravitational interaction.

\subsection{Sensitivity to $g_{p,e}\tilde{J}^{e,p}_Z$}
\begin{figure}
    \centering
    \includegraphics[width=1\linewidth]{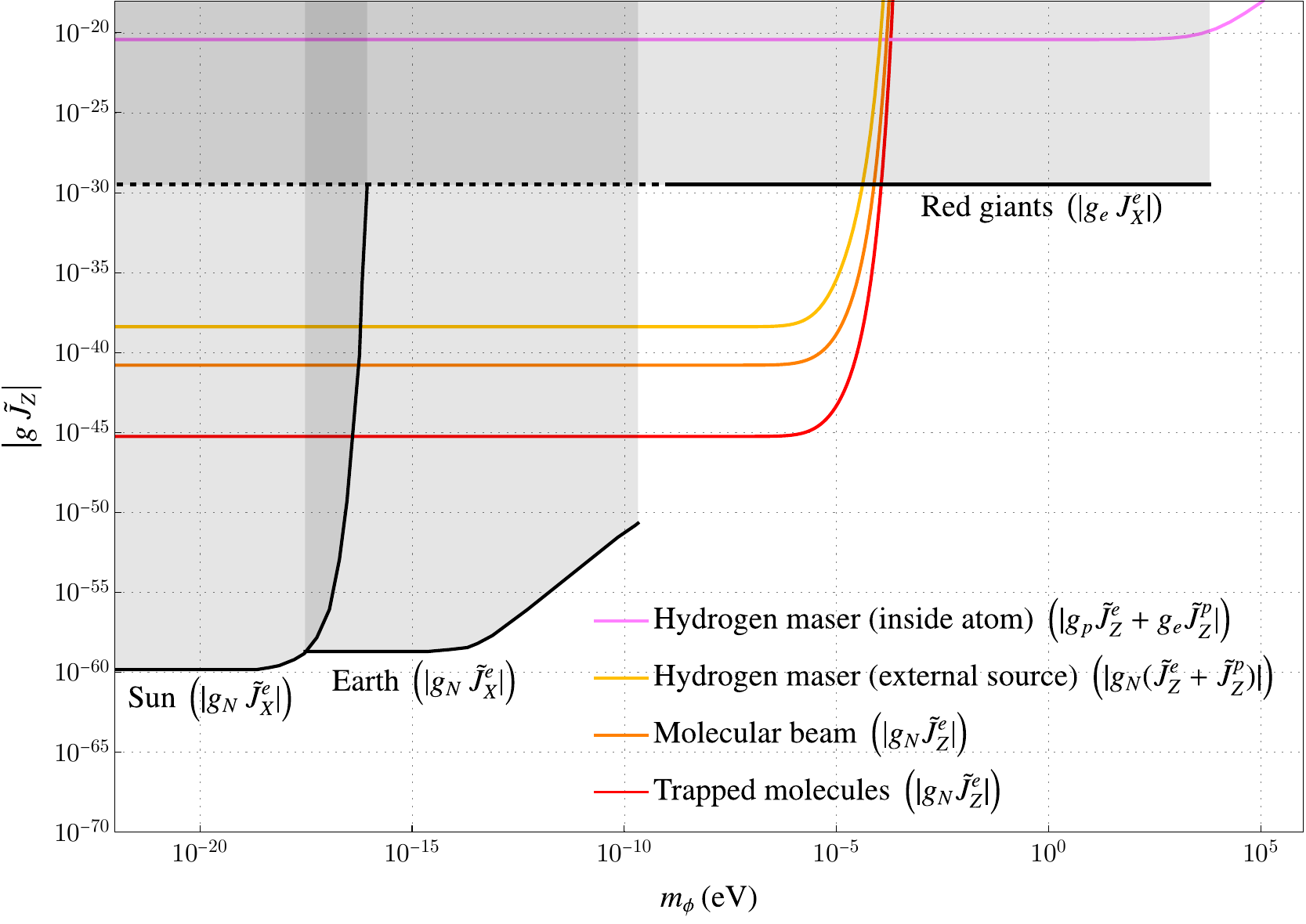}
    \caption{Summary of the projected sensitivities to Lorentz-violating coupling along Earth's rotation axis. The gray-shaded
    regions represent existing (although on different combination of couplings, discussed in the text) constraints \cite{Carenza:2025jwn}, while the unshaded regions with the magenta (hydrogen maser without external source, c.f. Sec.~\ref{sec2}), yellow (hydrogen maser with external source, c.f. Sec.~\ref{sec3}), orange (molecular beam, c.f. Sec.~\ref{molecular-beam}) and red (trapped molecules, c.f. Sec.~\ref{ultracold-section}) curves indicate parameter space that   can be explored by the proposed experiments discussed in this work.}
    \label{couplingz}
\end{figure}

In the following, we discuss possible ways to probe the Lorentz-violating coupling along the $Z$ axis of the Sun-centered frame. The relevant mixed coupling can be probed within the same setup either by reversing the quantization axis or rotating the quantization axis, which can be done with the $C-$ field.

\subsubsection{Reversing the quantization axis}
Equation~\eqref{eq:17} shows that the $X$ and $Y$ components produce the sidereal signal $\delta\nu^\mathrm{sid}_{\mathrm{Zeeman}}\propto \mathcal{C}_X\cos(\omega_\oplus t)+\mathcal{C}_Y\sin(\omega_\oplus t)$, whereas the $Z$ component produces the time-independent contribution 
\begin{equation}
2\pi\delta\nu^{(Z)}_{\mathrm{Zeeman}}=-\frac{\cos\chi}{4\pi a_H} \frac{\mathcal{C}_Z}{\Big(1+\frac{m_\phi a_H}{2}\Big)^2}.
\label{eq:22}
\end{equation}
A constant frequency offset is difficult to distinguish from an ordinary static magnetic field contribution. Consequently, the existing sidereal hydrogen maser search does not provide a similarly clean direct bound on $\mathcal{C}_Z$.

A dedicated experiment could overcome this limitation by reversing the quantization axis of the $C-$ field. Consider two measurements for opposite quantization directions $\hat{\bm{n}}$ and $-\hat{\bm{n}}$ while keeping the magnitude of the ordinary magnetic field unchanged. Therefore, for $\hat{\bm{n}}\to -\hat{\bm{n}}$, $\bm{\mathcal{\bm{C}}}.(\hat{\bm{n}})\to \bm{\mathcal{\bm{C}}}.(-\hat{\bm{n}})$. Hence, 
\begin{equation}
\delta\nu_{\mathrm{Zeeman}}(-\hat{\bm{n}}(t))=\frac{\mathcal{F}(m_\phi)}{2\pi}[\bm{\mathcal{\bm{C}}}\cdot (-\hat{\bm{n}}(t))]=-\frac{\mathcal{F}(m_\phi)}{2\pi}[\bm{\mathcal{\bm{C}}}\cdot (\hat{\bm{n}}(t))]=
-\delta\nu_{\mathrm{Zeeman}}(+\hat{\bm{n}}(t)).
\label{eq:23}
\end{equation}
By contrast, the leading ordinary Zeeman splitting is determined by the magnitude of the applied field $\Delta E_\mathrm{Zeeman}=g_F\mu_Bm_F|\bm{B}_C|$. Thus, reversing the field while keeping its magnitude fixed, $|\bm{B}_C|\to |-\bm{B}_C|=|\bm{B}_C|$, the leading ordinary Zeeman frequency remains approximately unchanged, i.e., $\nu^{(0)}_{\mathrm{Zeeman}}(+\hat{\bm{n}})\simeq \nu^{(0)}_{\mathrm{Zeeman}}(-\hat{\bm{n}})$. Thus, the reversal difference $(\nu_{\rm{Zeeman}}(\hat{\bm n})-
\nu_{\rm{Zeeman}}(-\hat{\bm n}))$ cancels the leading reversal-even Zeeman contribution while doubling the reversal-odd LV term. 

We write $\nu_\mathrm{Zeeman}(\hat{\bm{n}}(t))=\nu^{(0)}_\mathrm{Zeeman}+\delta\nu_{\mathrm{Zeeman}}(\hat{\bm{n}},t)$ and hence the reversal difference is 
\begin{equation}
\Delta\nu_{\mathrm{rev}}(t)=\nu_{\mathrm{Zeeman}}(\hat{\bm{n}}(t))-\nu_{\mathrm{Zeeman}}(-\hat{\bm{n}}(t))=2\delta\nu_{\rm{Zeeman}}(\hat{\bm{n}}(t)).   
\label{eq:24}
\end{equation}
For a quantization axis $\hat{\bm{n}}$, the Lorentz-violating contribution to the Zeeman frequency is 
\begin{equation}
2\pi \delta\nu_{\rm{Zeeman}}(\hat{\bm{n}}(t))=-\frac{1}{4\pi a_H}\frac{{\mathcal{C}_{\hat{n}}^{\rm{lab}}}(t)}{\Big(1+\frac{m_\phi a_H}{2}\Big)^2},
\label{eq:25}
\end{equation}
where in the case of laboratory $z$ axis $(\hat{n}=\hat{z})$, and Eq.~\eqref{eq:25} becomes
\begin{equation}
2\pi \delta\nu_{\rm{Zeeman}}(t)=-\frac{1}{4\pi a_H} \frac{\Big[\sin\chi \mathcal{C}_X\cos(\omega_\oplus t)+\sin\chi\mathcal{C}_Y\sin(\omega_\oplus t)+\cos\chi\mathcal{C}_Z\Big]}{\Big(1+\frac{m_\phi a_H}{2}\Big)^2}.  
\label{eq:26}
\end{equation}
Therefore, averaging over one sidereal day and comparing Eqs.~\eqref{eq:24} and \eqref{eq:25} results 
\begin{equation}
\langle2\pi \Delta\nu_{\rm{rev}}(t)\rangle=\mathcal{A}_E^{\rm{rev}}=-\frac{2}{4\pi a_H}\frac{\cos\chi \mathcal{C}_Z}{\Big(1+\frac{m_\phi a_H}{2}\Big)^2}.
\label{eq:27}
\end{equation}
Hence, a controlled field-reversal experiment with differential energy sensitivity $\mathcal{A}_E^{\rm{rev}}$ would probe
\begin{equation}
|\mathcal{C}_Z|=|g_p\tilde{J}^e_Z+g_e\tilde{J}^p_Z|\lesssim \frac{2\pi a_H \mathcal{A}^{\rm{rev}}_E}{\cos\chi} \Big(1+\frac{m_\phi a_H}{2}\Big)^2. 
\label{eq:28}
\end{equation}
If a future reversal measurement achieved a differential sensitivity comparable to $0.37~\mathrm{mHz}$, corresponds to $\mathcal{A}_E^{\rm{rev}}=1.53\times 10^{-27}~\mathrm{GeV}$, the sensitivity on the LV coupling (one coupling at a time) along the Sun-centered $Z$-axis could be obtained as 
\begin{equation}
|\mathcal{C}_Z|=|g_p\tilde{J}^e_Z|, |g_e\tilde{J}^p_Z| \sim 3.8\times 10^{-21}  \Big(1+\frac{m_\phi}{7.45~\mathrm{keV}}\Big)^2.
\label{eq:29}
\end{equation}

In FIG.~\eqref{couplingz}, we present the projected sensitivity to the Lorentz-violating coupling combination $|g_p\tilde{J}_Z^e+g_e\tilde{J}_Z^p|$ for scalar masses of $\mathcal{O}(\mathrm{keV})$, represented by the solid magenta curve.

\subsubsection{Rotating the quantization axis}
The constraints and sensitivity for the coupling obtained in the preceding discussion consider the quantization axis along the laboratory $z$ direction and hence, the Lorentz-violating coupling along the $z$ axis can be written in terms of the corresponding couplings in the Sun-centered frame, following Eq.~\eqref{eq:16}. Similarly, we can write laboratory frame Lorentz-violating vector components along $x$ and $y$ axis in terms of the components in the Sun-centered frame as
\begin{align}
 \tilde{J}_x(t)&=\cos\chi\Big[\tilde{J}_X\cos(\omega_\oplus t)+\tilde{J}_Y\sin(\omega_\oplus t)\Big]-\tilde{J}_Z\sin\chi,\nonumber\\
 \tilde{J}_y(t)&=-\tilde{J}_X\sin(\omega_\oplus t)+\tilde{J}_Y\cos(\omega_\oplus t).
 \label{eq:30}
\end{align}
If the quantization axis is fixed to the laboratory $x-$ axis by the $C-$ field, then we can write the LV shift as 
\begin{equation}
2\pi \delta\nu^x_{\rm{{Zeeman}}}(t)=-\frac{1}{4\pi a_H}  \frac{1}{\Big(1+\frac{m_\phi a_H}{2}\Big)^2} \Big[\cos\chi \mathcal{C}_X\cos(\omega_\oplus t)+\cos\chi\mathcal{C}_Y\sin(\omega_\oplus t)-\sin\chi\mathcal{C}_Z\Big], 
\label{eq:31}
\end{equation}
where the corresponding LV-induced Zeeman transition frequency for quantization axis along $\hat{z}$ is given in Eq.~\eqref{eq:17}. Comparing Eqs.~\eqref{eq:17} and \eqref{eq:31}, we obtain 
\begin{equation}
2\pi( \cos\chi \delta\nu^z_{\rm{Zeeman}}(t)-\sin\chi \delta\nu^x_{\rm{Zeeman}}(t))=2\pi \delta\nu_{\mathrm{rot}}(t)=-\frac{1}{4\pi a_H} \frac{\mathcal{C}_Z}{\Big(1+\frac{m_\phi a_H}{2}\Big)^2},  
\label{eq:32}
\end{equation}
where the experimentally measured quantity is $\delta\nu_{\mathrm{rot}}(t)=\cos\chi\delta\nu^z_{\rm{zeeman}}(t)-\sin\chi\delta\nu^x_{\mathrm{Zeeman}}(t)$. If a future rotational measurement achieved a differential sensitivity comparable to $0.37~\mathrm{mHz}$, correspond to $\mathcal{A}^{\rm{rot}}_E=1.53\times 10^{-27}~\mathrm{GeV}$, the sensitivity on the LV coupling (one coupling at a time) along the Sun-centered $Z$-axis could be obtained as 
\begin{equation}
|\mathcal{C}_Z|=|g_p\tilde{J}^e_Z|, |g_e\tilde{J}^p_Z| \sim5.2\times 10^{-21} \Big(1+\frac{m_\phi}{7.45~\mathrm{keV}}\Big)^2.  
\label{eq:33}
\end{equation}

\section{Hydrogen-maser probe of scalar-mediated Lorentz violation in presence of an external source}\label{sec3}

\begin{figure}
    \centering
    \includegraphics[width=12cm]{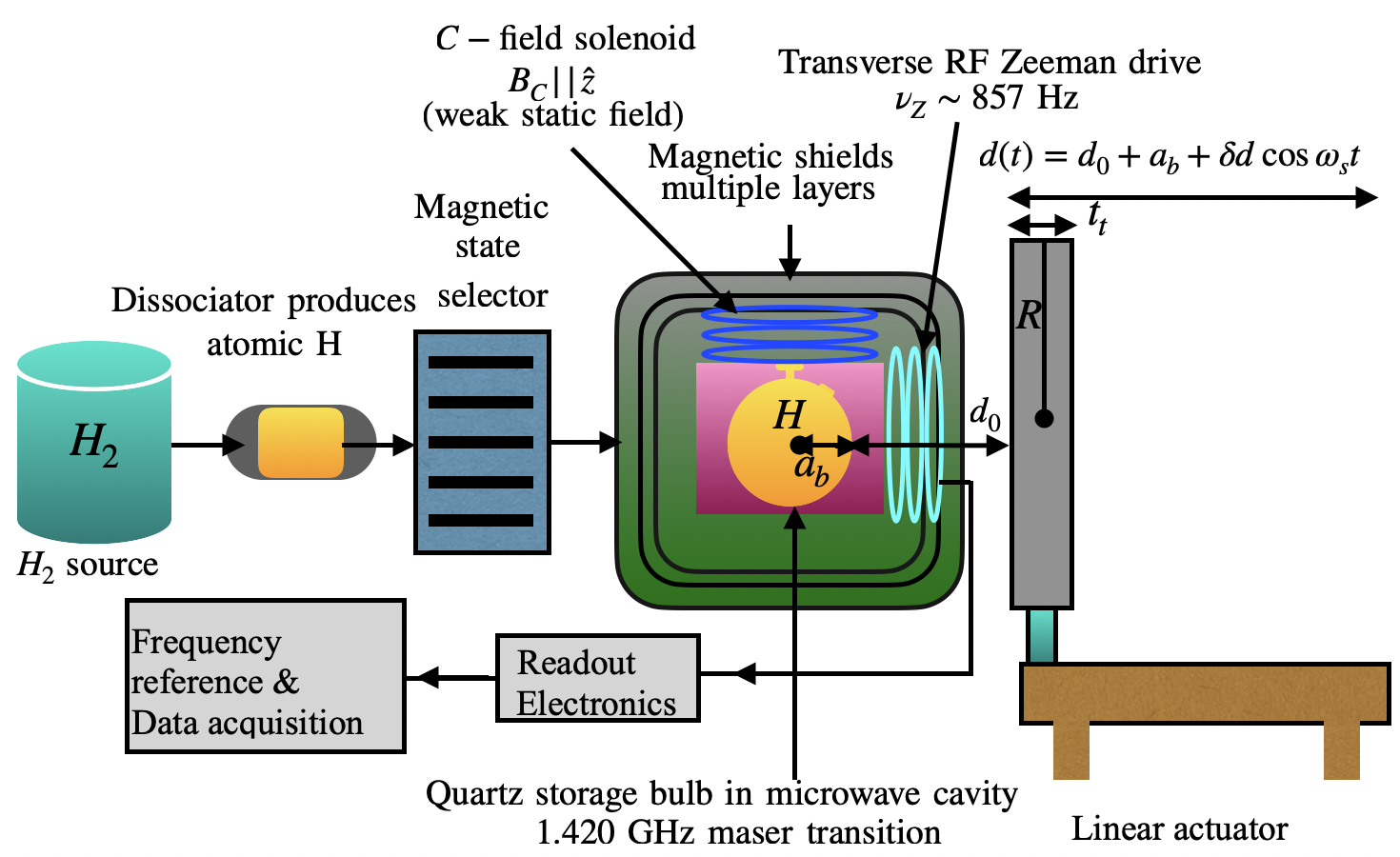}
    \caption{Hydrogen maser apparatus with a moving source.
    }
    \label{fig3}
\end{figure}

While the previous section explored the sensitivity of scalar-mediated Lorentz-violating coupling using hydrogen maser spectroscopy to scalar masses of $\mathcal{O}(\mathrm{keV})$, in this section we extend the analysis to substantially lighter scalar mediators. In the case of the $\mathcal{O}(\mathrm{keV})$ scalar mediator, the corresponding interaction occurs between the electron and proton within a hydrogen atom, and the characteristic Bohr radius therefore sets the relevant interaction length scale and, consequently, the accessible scalar mass range. Here, we propose a complementary configuration in which the scalar field is sourced by the nucleons of an external macroscopic object. The resulting scalar field can interact with the electron and proton spins in hydrogen atoms and induce shifts in the maser energy levels, particularly in the Zeeman transitions. Since the characteristic source-sensor separation can be much larger than the Bohr radius, this setup extends the sensitivity of hydrogen maser spectroscopy toward significantly lighter scalar masses.

In the proposed experimental configuration illustrated in FIG.~\eqref{fig3}, a dense cylindrical source plate mounted on a linear actuator is positioned outside the magnetic shielding of the conventional hydrogen maser apparatus shown in FIG.~\eqref{fig0}. The plate is translated periodically toward and away from the maser so that its distance from the hydrogen sample varies sinusoidally. The macroscopic source object generates the scalar field through its ordinary scalar coupling to matter nucleons, while the scalar interacts with the electron and/or proton spin of the hydrogen atom through the Lorentz-violating coupling. Consequently, the source-induced contribution to the hydrogen Zeeman splitting is periodically modulated at $\omega_s$. The interplay between the periodic source modulation and the Earth's rotation relative to the preferred Lorentz-violating background gives rise to characteristic spectral signatures at the source modulation frequency $\omega_s$ and the sidereal sidebands $\omega_s\pm\omega_\oplus$.

An important advantage of periodically modulating the external source is that the resulting signal can be distinguished from contributions to the scalar field generated by stationary matter surrounding the maser, including the apparatus itself. While such static sources can produce sidereal variations of the Zeeman frequency, they do not generate signals at the source-modulation frequency $\omega_s$ or its sidereal sidebands $\omega_s\pm\omega_\oplus$, allowing the controlled source contribution to be isolated. Such periodic modulation of the external source can also help distinguish the scalar-mediated interaction from potential gravitational interactions. 

\begin{figure}
    \centering
    \includegraphics[width=12cm]{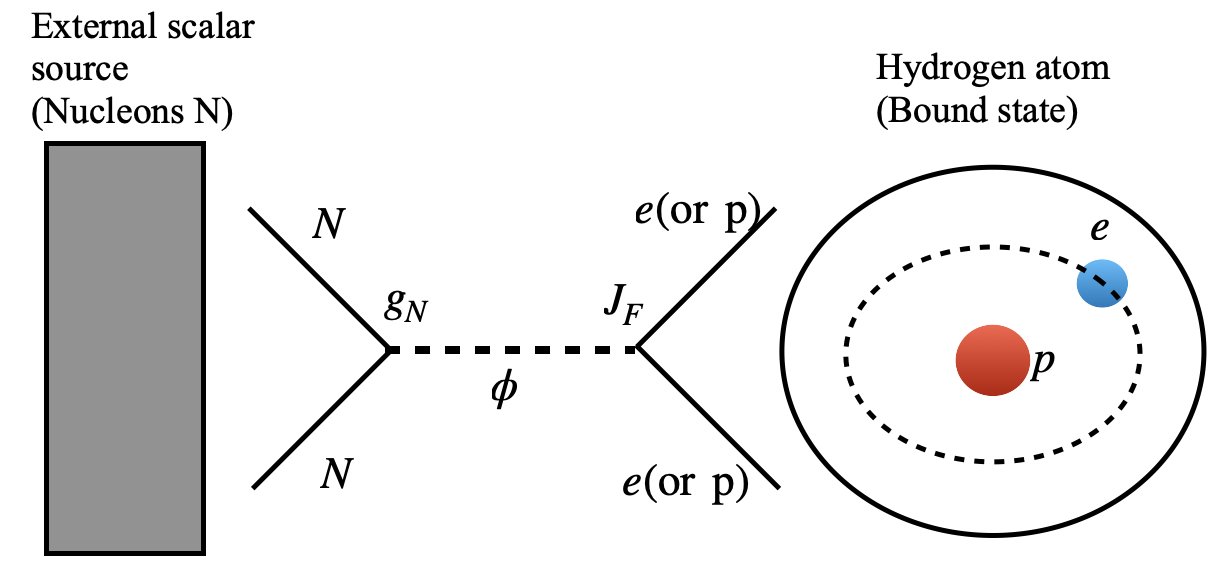}
    \caption{Feynman diagram for a Lorentz-violating scalar-mediated interaction.}
    \label{fig4}
\end{figure}

We place a macroscopic laboratory source at a characteristic distance of $\mathcal{O}(\mathrm{cm})$ from the hydrogen maser detector. Such a source-detector separation provides sensitivity to scalar masses around the $\mathcal{O}(\mu\mathrm{eV})$ scale, since the range of the Yukawa interaction is set by $m_\phi^{-1}$. More generally, the experiment is most sensitive when the Compton wavelength of the scalar is comparable to, or larger than, the characteristic separation between the source and the hydrogen atoms. We consider the Lorentz-invariant scalar coupling to the unpolarized laboratory source given by the Lagrangian
\begin{equation}
\mathcal{L}_\mathrm{source}=g_N\phi \bar{N}N,
\label{eq:a1}
\end{equation}
where $N$ denotes the source nucleon. At the hydrogen detector, we consider either electron, or proton or both have Lorentz-violating couplings given by the Lagrangian
\begin{equation}
\mathcal{L}_\mathrm{LV}=-\phi\sum _{F=e,p} \bar{\psi}_F\Big[J^F_\mu\gamma_5\gamma^\mu+\frac{1}{2}L^F_{\mu\nu}\sigma^{\mu\nu}\Big]\psi_F, 
\label{eq:a2}
\end{equation}
where the corresponding Feynman diagram is shown in FIG.~\eqref{fig4}. In the non-relativistic limit of the fermions, we can write the potential between two point-like fermions separated by a distance $r$ as
\begin{equation}
V_{\mathrm{LV}}(r) =g_N(\bm{\sigma}_e\cdot \tilde{\bm{J}}^e+\bm{\sigma}_p\cdot \tilde{\bm{J}}^p)f(r),
\label{eq:a3}
\end{equation}
where $\tilde{J}$ and $f(r)$ are already defined. For an extended macroscopic source composed of unpolarized nucleons, the pointlike Yukawa potential must be generalized to account for the finite spatial distribution of the source. Treating the hydrogen maser apparatus as approximately point-like relative to the source dimensions and source-detector separation, the resulting interaction can be obtained by integrating the Yukawa contribution over the nucleon density of the source as
\begin{equation}
f(r)\to I_s(\mathbf{r})=\int d^3r^\prime n_N(\bm{r}^\prime)\frac{e^{-m_\phi|\bm{r}-\bm{r}^\prime|}}{4\pi|\bm{r}-\bm{r}^\prime|},
\label{eq:a4}
\end{equation}

where $n_N$ is the nucleon number density. To evaluate Eq.~\eqref{eq:a4}, and hence obtain the LV-induced Zeeman transition frequency, one must specify the geometry of the source. The source can be of a uniform sphere or a finite cylindrical plate. We choose the flat plate because of the advantage in geometric efficiency. For a flat plate facing the detector, a substantial fraction of the source material can remain close to the minimum separation. In contrast, the curved surface such as a sphere causes most of its volume to lie at larger distances from the detector. As a result, for a fixed source mass and closest approach, a plate geometry can generally yield a stronger short-range signal. \footnote{The source integral in Eq.~\eqref{eq:a4} are different for different source geometries and are discussed in Appendix.~\eqref{appendix1}.}

The external scalar field is essentially constant across one hydrogen atom provided $d\gg a_H$. Thus, we can write the perturbing Hamiltonian operator 
\begin{equation}
\delta H_{\mathrm{LV}}=g_N I_s\Big[\bm{\sigma}_e\cdot \tilde{\bm{J}}^e+\bm{\sigma}_p\cdot \tilde{\bm{J}}^p\Big]. 
\label{eq:a5}
\end{equation}
Considering the laboratory $z$-axis as the spin quantization axis, defined by the applied static $C$-field, the first-order Lorentz-violating energy shifts are obtained as
\begin{align}
\delta E_{+1}&=I_s(g_N\tilde{J}_z^e+g_N \tilde{J}^p_z),\nonumber\\
\delta E_0&=0,\nonumber\\
\delta E_{-1}&=-I_s(g_N\tilde{J}_z^e+g_N\tilde{J}^p_z).
\label{eq:a6}
\end{align}

The maser experiment gives the averaged $F=1$ Zeeman frequency given in Eq.~\eqref{eq:12} and hence we can write the correction to the Zeeman frequency due to LV as
\begin{equation}
\delta\nu_{\rm{Zeeman}}=    \frac{I_sg_N(\tilde{J}_z^e+\tilde{J}^p_z)}{2\pi}.
\label{eq:a7}
\end{equation}
Experimentally, the Zeeman transition frequency $\nu^{\rm{meas}}_Z(t)$ needs to be searched over time for any sidereal modulation caused by the LV.
We define $\tilde{J}_z^e+\tilde{J}^p_z=\tilde{J}_z$, which are functions of time and can be transformed into the Sun-centered inertial frame as  
\begin{equation}
\tilde{J}_z(t)=\sin\chi\Big[\tilde{J}_X\cos(\omega_\oplus t)+\tilde{J}_Y\sin(\omega_\oplus t)\Big]+\tilde{J}_Z\cos\chi, 
\label{eq:a8}
\end{equation}
and using Eqs.~\eqref{eq:a7} and \eqref{eq:a8}, we obtain 
\begin{equation}
2\pi \delta\nu_{\rm{Zeeman}}(t)=g_NI_s\Big[\sin\chi\Big(\tilde{J}_X\cos(\omega_\oplus t)+\tilde{J}_Y\sin(\omega_\oplus t)\Big)+\tilde{J}_Z\cos\chi\Big].   
\label{eq:a9}
\end{equation}
Thus, the transverse part gives a sidereal amplitude 
\begin{equation}
\mathcal{A}_\perp=g_NI_s\sin\chi J_\perp,~~J_\perp=\sqrt{\tilde{J^2_X}+\tilde{J^2_Y}}.
\label{eq:a10}
\end{equation}
However, for a stationary source, isolating a Lorentz-violating signal and constraining the corresponding coupling can be challenging, since the effect may be indistinguishable from that of other conventional static scalar sources in the background. Also, the $\tilde{J}_Z$ contribution would appear only as a constant shift of the Zeeman frequency. Thus, a movable source can be a useful discriminator compared to any static scalar source background.

Suppose the plate physically oscillates toward and away from the hydrogen maser using a linear actuator and the source-detector separation varies sinusoidally as
\begin{equation}
d(t)=d_0+\delta d\cos\omega_s t,
\label{eq:a11}
\end{equation}
where $d_0$ is the mean distance, $\delta d$ is the displacement amplitude and $\omega_s$ is the mechanical modulation frequency. Accordingly, we can write the source integral as
\begin{equation}
I_s(t)=I_0+I_1\cos\omega_s t+\cdots, 
\label{eq:a12}
\end{equation}
where $I_0$ represents the time-averaged DC component of the source integral, while $I_1$ denotes the first-harmonic amplitude at the source modulation frequency $\omega_s$. The ellipses account for contributions from higher harmonics. For the present analysis, we retain only the first harmonic, which is sufficient to characterize the leading sidereal sidebands at $\omega_s\pm\omega_\oplus$.
Therefore, Eq.~\eqref{eq:a9} becomes
\begin{align}
2\pi\,\delta\nu_{\rm Zeeman}(t)
={}&
g_N I_0 \widetilde{J}_Z \cos\chi
+ g_N I_1 \widetilde{J}_Z \cos\chi \cos(\omega_s t)
\nonumber\\
&+ g_N I_0 \sin\chi\,\widetilde{J}_X \cos(\omega_\oplus t)
+ g_N I_0 \sin\chi\,\widetilde{J}_Y \sin(\omega_\oplus t)
\nonumber\\
&+ \frac{1}{2} g_N I_1 \widetilde{J}_X \sin\chi
\left[
\cos\bigl((\omega_s+\omega_\oplus)t\bigr)
+
\cos\bigl((\omega_s-\omega_\oplus)t\bigr)
\right]
\nonumber\\
&+ \frac{1}{2} g_N I_1 \widetilde{J}_Y \sin\chi
\left[
\sin\bigl((\omega_s+\omega_\oplus)t\bigr)
-
\sin\bigl((\omega_s-\omega_\oplus)t\bigr)
\right],
\label{eq:a13}
\end{align}
where the first time-independent DC term acts as a constant offset in the Zeeman transition frequency and does not contribute to the time-modulated signal. Thus, the signal contains frequencies $\omega_s$ associated with the search for Lorentz-violating $Z$ component and $\omega_s\pm\omega_\oplus$ associated with the search for Lorentz-violating $X$ and $Y$ components. The pure $\omega_\oplus$ signal is at very low frequency and is mostly affected by the drift and $1/f$ noise. Thus, the clean high frequency signals are the carrier frequency $\omega_s$ and the two sideband frequencies $\omega_s\pm \omega_\oplus$, which are sensitive to $\tilde{J}_Z$ with amplitude $g_N I_1\cos\chi\tilde{J}_Z$ and $\tilde{J}_X, \tilde{J}_Y$ with amplitudes $(1/2)g_NI_1\sin\chi\tilde{J}_X, (1/2)g_NI_1\sin\chi\tilde{J}_Y$ respectively \footnote{Alternatively, instead of sinusoidally modulating the source-detector separation, one may periodically switch the scalar source between near and far positions, as discussed in Appendix.~\eqref{appendix2}.}.

To visualize the Lorentz-violating signal, characterizing the noise background, and search for excess spectral power at the predicted signal frequencies, it is useful to examine the corresponding power spectral density (PSD). Thus, we can calculate the Fourier transform of the Lorentz-violating Zeeman transition frequency in the frequency space as
\begin{equation}
 \tilde{\delta\nu}_T(\omega)=\int ^T_0 dt e^{i\omega t} \delta\nu_{\mathrm{Zeeman}}(t),   
 \label{eq:a14}
\end{equation}
over an observation time $T$. Therefore, near each positive frequency peak, the Fourier amplitudes are
\begin{equation}
\begin{split}
\tilde{\delta\nu}_T(\omega\simeq\omega_s)&= \frac{1}{4\pi}g_NI_1\tilde{J}_Z\cos\chi\mathcal{I}_T(\omega-\omega_s),\\
\tilde{\delta\nu}_T(\omega\simeq\omega_s-\omega_\oplus)&=\frac{1}{8\pi}g_NI_1\sin\chi(\tilde{J}_X-i\tilde{J}_Y)\mathcal{I}_T(\omega-(\omega_s-\omega_\oplus)),\\
\tilde{\delta\nu}_T(\omega\simeq\omega_s+\omega_\oplus)&=\frac{1}{8\pi}g_NI_1\sin\chi(\tilde{J}_X+i\tilde{J}_Y)\mathcal{I}_T(\omega-(\omega_s+\omega_\oplus)),
\end{split}
\label{eq:a15}
\end{equation}
up to the negative frequency mirror peaks, where
\begin{eqnarray}
\mathcal{I}_T(\omega-\omega_i)&=&\int ^T_0 dt e^{i(\omega-\omega_i)t}\nonumber\\
&=&Te^{i(\omega-\omega_i)T/2}\operatorname{sinc}\Big[\frac{(\omega-\omega_i)T}{2}\Big],    
\label{eq:a16}
\end{eqnarray}
with the definition $\operatorname{sinc}x=\sin x/x$.

Thus, the finite time Fourier power spectrum is obtained as $P_T(\omega)=|\tilde{\delta\nu}_T(\omega)|^2$, where 
\begin{equation}
\begin{split}
P_T(\omega\simeq\omega_s)&=\Big(\frac{g_NI_1\tilde{J}_Z}{4\pi}\Big)^2\cos^2\chi T^2\operatorname{sinc}^2\Big[\frac{(\omega-\omega_s)T}{2}\Big],\\
P_T(\omega\simeq\omega_s-\omega_\oplus)&=\Big(\frac{g_NI_1}{8\pi}\Big)^2(\tilde{J^2_X}+\tilde{J^2_Y})\sin^2\chi T^2\operatorname{sinc}^2\Big[\frac{(\omega-\omega_s+\omega_\oplus)T}{2}\Big],\\
P_T(\omega\simeq\omega_s+\omega_\oplus)&=\Big(\frac{g_NI_1}{8\pi}\Big)^2(\tilde{J^2_X}+\tilde{J^2_Y})\sin^2\chi T^2\operatorname{sinc}^2\Big[\frac{(\omega-\omega_s-\omega_\oplus)T}{2}\Big].
\label{eq:a17}
\end{split}
\end{equation}

For a real-valued time series, the one-sided PSD for $f>0$ is 
\begin{equation}
S_\nu(f)=\frac{2}{T}|\tilde{\delta\nu}_T(f)|^2=\frac{2}{T}P_T(f),   \label{eq:a18} 
\end{equation}
and the corresponding amplitude spectral density is $\sqrt{S_\nu(f)}$ in the unit of $\mathrm{Hz}/\sqrt{\mathrm{Hz}}$. For the readout, we are not exactly Fourier transforming the $857~\mathrm{Hz}$ Zeeman oscillation itself in this analysis. Rather, the experiment would repeatedly determines the Zeeman resonance frequency with a slowly varying $\nu_Z(t)$ and we calculate the PSD of the variations of that measured resonance frequency. Thus, we obtain
\begin{equation}
\begin{split}
S_\nu(f\simeq f_s)&=\frac{T}{2}    \Big(\frac{g_NI_1\tilde{J}_Z}{2\pi}\Big)^2\cos^2\chi \operatorname{sinc}^2\Big[\pi(f-f_s)T\Big],\\
S_\nu(f\simeq f_\pm)&=\frac{T}{2}\Big(\frac{g_NI_1}{4\pi}\Big)^2(\tilde{J^2_X}+\tilde{J^2_Y})\sin^2\chi \operatorname{sinc}^2\Big[\pi(f-f_\pm)T\Big],~~~f_\pm=f_s\pm f_\oplus.
\label{eq:a19}
\end{split}
\end{equation}
The finite observation time  broadens each monochromatic contribution into a $\sin c^2$ profile with a characteristic width $\Delta f\sim 1/T$. The Lorentz-violating interaction can therefore be searched for as excess spectral power at the controlled source-modulation frequency $f_s$ and the sidebands $f_s\pm f_\oplus$. The heights of the peaks of $\sqrt{S_\nu(f)}$ determine the strength of the Lorentz-violating coefficients. To resolve the sidereal structure, the integration time should be taken as $T\gtrsim 1/f_\oplus$.

\subsection{Effect of the finite size detector}
The above expressions given in Eqs.~\eqref{eq:a15} and \eqref{eq:a19} for the LV-induced Zeeman transition frequency and hence the PSD are calculated with the assumption of point-detector source integral. These equations correspond to the scalar field evaluated at a single observation point, which we take to be the center of the hydrogen storage bulb. In practice, however, the hydrogen atoms occupy a finite volume $V_b=4\pi a^3_b/3$, where $a_b$ is the bulb radius. We therefore average the source-induced Yukawa field over the storage bulb. Thus, for the finite cylinder plate, the source integral Eq.~\eqref{eq:a4} becomes \footnote{The detail derivation of the source integral for a finite size detector hydrogen bulb is given in Appendix.~\eqref{appendix3}.}
\begin{equation}
\bar{I}_s(d;m_\phi)=\mathcal{F}_b(m_\phi a_b)\frac{n_N}{2m_\phi}\int_{a_b+d}^{a_b+d+t_t}  dz\Big[e^{-m_\phi z}-e^{-m_\phi \sqrt{z^2+R^2}}\Big]=\mathcal{F}_b(m_\phi a_b)I_s^c(a_b+d;m_\phi),
\label{eq:a31}
\end{equation}
where 
\begin{equation}
\mathcal{F}_b(m_\phi a_b)=\frac{3[m_\phi a_b\cosh(m_\phi a_b)-\sinh(m_\phi a_b)]}{(m_\phi a_b)^3}.  
\label{eq:a32}
\end{equation}
Therefore, the sinusiodal variation of the distance from the bulb center to the nearest plate surface can be written from Eq.~\eqref{eq:a11} as
\begin{equation}
d(t)=d_0+a_b+\delta d\cos\omega_s t,    
\end{equation}
where $d_0$ is now the surface to surface distance between the source and the bulb. The point detector source integral at the bulb center is therefore 
\begin{equation}
I_s^c(t)=I_s[d(t)]\simeq I_0^c+I_1^c \cos\omega_s t+\cdots,   
\end{equation}
where
\begin{equation}
I_1^c=\frac{1}{\pi}\int ^{2\pi}_0 d\theta I_s(a_b+d_0+\delta d\cos\theta)\cos\theta,  ~~\theta=\omega_s t 
\label{int1}
\end{equation}
and 
\begin{equation}
I_s(d(\theta);m_\phi)=\frac{n_N}{2m_\phi}\int^{d(\theta)+t_t}_{d(\theta)}  dz\Big[e^{-m_\phi z}-e^{-m_\phi \sqrt{z^2+R^2}}\Big].
\label{int2}
\end{equation}

After averaging over the spherical hydrogen bulb, we obtain
\begin{equation}
\bar{I}_s(t)= \mathcal{F}_b(m_\phi a_b) I_s^c(t).   
\end{equation}
Therefore, $I_1$ in Eqs.~\eqref{eq:a13}, \eqref{eq:a15}, \eqref{eq:a17} and \eqref{eq:a19} will be replaced by $\bar{I}_1=\mathcal{F}_b(m_\phi a_b) I_1^c$ for a finite size detector, which needs to be calculated numerically for a finite cylindrical plate with non-zero scalar mass.

\subsection{Sensitivity to the Lorentz-violating coupling}
In this section, we obtain sensitivities on the Lorentz-violating couplings in the Sun-centered frame using the hydrogen maser spectroscopy. The Earth's sidereal frequency is
\begin{equation}
f_\oplus=\frac{1}{T_\oplus}\simeq \frac{1}{23~\mathrm{h}~56~\mathrm{min}}\simeq 11.6~\mathrm{\mu\mathrm{Hz}}.   
\label{eq:a20}
\end{equation}
To adequately characterize the associated sidereal modulation, we adopt a total observation time of $T=10^6~\mathrm{s}\simeq11.6~\mathrm{days}$. This choice provides a sufficiently long data set to resolve the characteristic time-dependent structure of the Lorentz-violating signal and to perform a spectral analysis of its modulation components.

For the sensitivity estimate of the Lorentz-violating coupling, we adopt a source-modulation frequency $f_s=100~\mu{\rm Hz}$, corresponding to a modulation period
\begin{equation}
T_s=f_s^{-1}=10^4~{\rm s}\simeq 2.78~{\rm h}.
\label{eq:a21}
\end{equation}
For an observation time $T\simeq11.6~{\rm days}$, the data therefore span approximately $N_s=T/T_s=100$ complete source-modulation cycles. Over the same interval, the number of sidereal cycles is
\begin{equation}
N_\oplus=\frac{T}{T_\oplus}\simeq 11.6,
\label{eq:a22}
\end{equation}
so that both the source modulation and the sidereal variation are sampled over many periods. The carrier occurs at $f_s=100~\mu{\rm Hz}$, while the Lorentz-violating sidereal sidebands are located at
$f_-=f_s-f_\oplus\simeq88.4~\mu{\rm Hz}$ and $f_+=f_s+f_\oplus\simeq111.6~\mu{\rm Hz}$, where $f_\oplus\simeq11.6~\mu{\rm Hz}$. For $T=10^6~{\rm s}$, the Fourier-frequency resolution is $\Delta f=1/T=1~\mu{\rm Hz}$. Consequently, each sideband is separated from the carrier by $f_\oplus/\Delta f\simeq11.6$ Fourier bins, while the two sidebands are separated from one another by $2f_\oplus/\Delta f\simeq23.2$ bins. They are therefore well resolved for the assumed observation time.

Taking a representative sampling interval for the Zeeman frequency measurement of $t_r=18~{\rm min}=1080~{\rm s}$ \cite{Humphrey:2001wm}, each source modulation cycle contains approximately
\begin{equation}
\frac{T_s}{t_r}\simeq9.3
\label{eq:a23}
\end{equation}
measurements, while the full data set contains approximately $T/t_r\simeq926$ samples. The corresponding sampling frequency is
\begin{equation}
f_{\rm samp}=t_r^{-1}\simeq926~\mu{\rm Hz},
\label{eq:a24}
\end{equation}
and hence the Nyquist frequency (highest frequency that can be measured) is
\begin{equation}
f_{\rm Nyq}=\frac{f_{\rm samp}}{2}\simeq463~\mu{\rm Hz}.
\label{eq:a25}
\end{equation}
Since $f_+\simeq111.6~\mu{\rm Hz}\ll f_{\rm Nyq}$, the carrier and both sidereal sidebands lie comfortably within the measurable frequency range. Thus, the chosen cadence (refers to the time interval between consecutive measurements) is sufficient to sample the relevant oscillatory components without aliasing (occurs when a high frequency signal is sampled too slowly and consequently appears as a false, lower frequency signal), while the total observation time provides adequate spectral resolution to distinguish the carrier from the sidereal sidebands. It is useful to emphasize that these two quantities characterize different aspects of the measurement; $f_{\rm Nyq}$ sets the highest frequency that can be reconstructed unambiguously from the sampled data, whereas $\Delta f=1/T$ determines the ability to resolve nearby spectral features.

For the source trajectory, we adopt the benchmark parameters $d_0=10~\mathrm{cm}$ and $\delta d=5~\mathrm{cm}$, $a_b=8.85~\mathrm{cm}$ for $V_b=2.9\times 10^{-3}~\mathrm{m^{3}}$ \cite{Humphrey:2001wm} such that $d(t)=a_b+d_0+\delta d\cos\omega_s t$. The corresponding source velocity is
\begin{equation}
\dot d(t)=-\omega_s\delta d\sin\omega_s t,
\label{eq:a26}
\end{equation}
with a maximum value
\begin{equation}
v_{\rm max}=\omega_s\delta d\simeq3.14\times10^{-5}~\mathrm{m\,s^{-1}}
\simeq31~\mu\mathrm{m\,s^{-1}}.
\label{eq:a27}
\end{equation}
The maximum acceleration is similarly
\begin{equation}
a_{\rm max}=\omega_s^2\delta d\simeq2.0\times10^{-8}~\mathrm{m\,s^{-2}}.
\label{eq:a28}
\end{equation}
The source motion is therefore highly non-relativistic and adiabatically slow, justifying a quasi-static treatment in which, at each instant, the source may be approximated as a stationary plate located at the instantaneous separation $d(t)$.

During one such $t_r$ measurement interval, the phase of the source modulation changes by
\begin{equation}
\Delta\phi_s=\omega_s t_r
=2\pi f_s t_r
\simeq0.68~\mathrm{rad}
\simeq39^\circ.
\label{eq:a29}
\end{equation}
Thus, although the source advances by a non-negligible fraction of its modulation cycle during an individual frequency measurement, the finite integration time produces only a small attenuation of the measured modulation amplitude. Approximating each frequency determination as a uniform time average over the interval $t_r$, the carrier amplitude is reduced by the factor
\begin{equation}
\eta_s=
\frac{\sin(\pi f_s t_r)}
{\pi f_s t_r}
=\operatorname{sinc}(\pi f_s t_r)
\simeq0.981,
\label{eq:a30}
\end{equation}
corresponding to an amplitude suppression of only about $1.9\%$. At the upper sidereal sideband, $f_+\simeq111.6~\mu\mathrm{Hz}$, the corresponding averaging factor is $\eta_+\simeq0.976$, giving an attenuation of approximately $2.4\%$. Therefore, finite time averaging over an $18$ minutes Zeeman frequency measurement has only a minor effect on the carrier and sidereal sideband amplitudes, and the relevant modulation structure remains essentially preserved.

Consider a source tungsten plate of mass $M_s=100~\mathrm{kg}$, radius $R=25~\mathrm{cm}$, thickness $t_t=M_s/\rho_s\pi R^2\simeq 2.65~\mathrm{cm}$ for the tungsten density $\rho_s=1.925\times 10^4~\mathrm{kg}/\mathrm{m}^3$. For the previous choice of $d_0=10~\mathrm{cm}$, $\delta d=5~\mathrm{cm}$, $\chi=48^\circ$, $a_b=8.85~\mathrm{cm}$, $A_\nu^{\rm{exp}}\simeq 0.37~\mathrm{mHz}$, we have
\begin{equation}
d(t)=18.85~\mathrm{cm}+5~\mathrm{cm}\cos[2\pi (100~\mathrm{\mu Hz})t].    
\label{eq:a31}
\end{equation}
Therefore, the distance between the source surface and the bulb center varies as $13.85~\mathrm{cm}\leq d(t)\leq 23.85~\mathrm{cm}$. The finite cylinder source integral is $\bar{I}_1(m_\phi)=\mathcal{F}(m_\phi a_b)I_1^c(m_\phi)$, where $I_1^c(m_\phi)$ can be obtained from Eqs.~\eqref{int1} and \eqref{int2}.

In the long-range limit $\mathcal{F}(m_\phi a_b)\to 1$ as $m_\phi\to 0$ and we obtain  $|\bar{I}_1(0)|\simeq 5.7\times 10^{11}~\mathrm{GeV}$, numerically. Therefore, using Eq.~\eqref{eq:a13}, we obtain the sensitivity on the Lorentz-violating coupling as
\begin{equation}
|g_N\tilde{J}_Z|\sim \frac{2\pi A_\nu^{\mathrm{exp}}}{|\bar{I}_1||\cos\chi|} \sim 4.02\times 10^{-39}.   
\label{limit1}
\end{equation}
Similarly, for the transverse components
\begin{equation}
 |g_N\tilde{J}_\perp|\sim  \frac{4\pi A_\nu^{\mathrm{exp}}}{|\bar{I}_1||\sin\chi|} \sim  7.24\times 10^{-39},~~\tilde{J}_\perp=\sqrt{\tilde{J}_X^2+\tilde{J}_Y^2}. 
 \label{limit2}
\end{equation}
However, since each Zeeman frequency lasts $18~\mathrm{min}$, the sinusoidal variation changes the signal slightly and the measured amplitude reduced by a $\operatorname{sinc}$ factor. Therefore, 
\begin{equation}
\begin{split}
|g_N\tilde{J}_Z|_s&= \frac{|g_N\tilde{J}_Z|}{\eta_s}=\frac{|g_N(\tilde{J}^p_Z+\tilde{J}^e_Z)|}{\eta_s}\sim 4.10\times 10^{-39}, \\  |g_N\tilde{J}_\perp|_-&=\frac{ |g_N\tilde{J}_\perp|}{\eta_-}=\frac{ |g_N(\tilde{J}^p_\perp+\tilde{J}^e_\perp)|}{\eta_-}\sim 7.35\times 10^{-39},\\  |g_N\tilde{J}_\perp|_+&=\frac{|g_N\tilde{J}_\perp|}{\eta_+}=\frac{|g_N(\tilde{J}^p_\perp+\tilde{J}^e_\perp)|}{\eta_+}\sim 7.42\times 10^{-39}, 
\end{split}
\label{limit3}
\end{equation}
where $\eta_s\simeq 0.981$, $\eta_-\simeq 0.985$, $\eta_+\simeq 0.976$.

The above limits given in Eqs.~\eqref{limit1}, \eqref{limit2} and \eqref{limit3} are valid for zero scalar mass limit. In particular, the sensitivities given in Eq.~\eqref{limit1} and \eqref{limit2} are obtained before scan averaging, meaning the Zeeman frequency measurement is instantaneous, whereas the later sensitivities given in Eq.~\eqref{limit3} are obtained when the Zeeman frequency is measured by averaging the oscillatory signal over $18~\mathrm{min}$.

The characteristic scalar mass reach is set by the inverse of the maximum source-detector separation, $d_\mathrm{max}=d_0+\delta d+a_b=23.85~\mathrm{cm}$, corresponding to $m_\phi\simeq 8.3\times 10^{-7}~\mathrm{eV}$. For $m_\phi\ll 8.3\times 10^{-7}~\mathrm{eV}$, the interaction approaches the long-range regime and the sensitivity becomes approximately independent of the scalar mass.

\begin{figure}
    \centering
    \includegraphics[width=12cm]{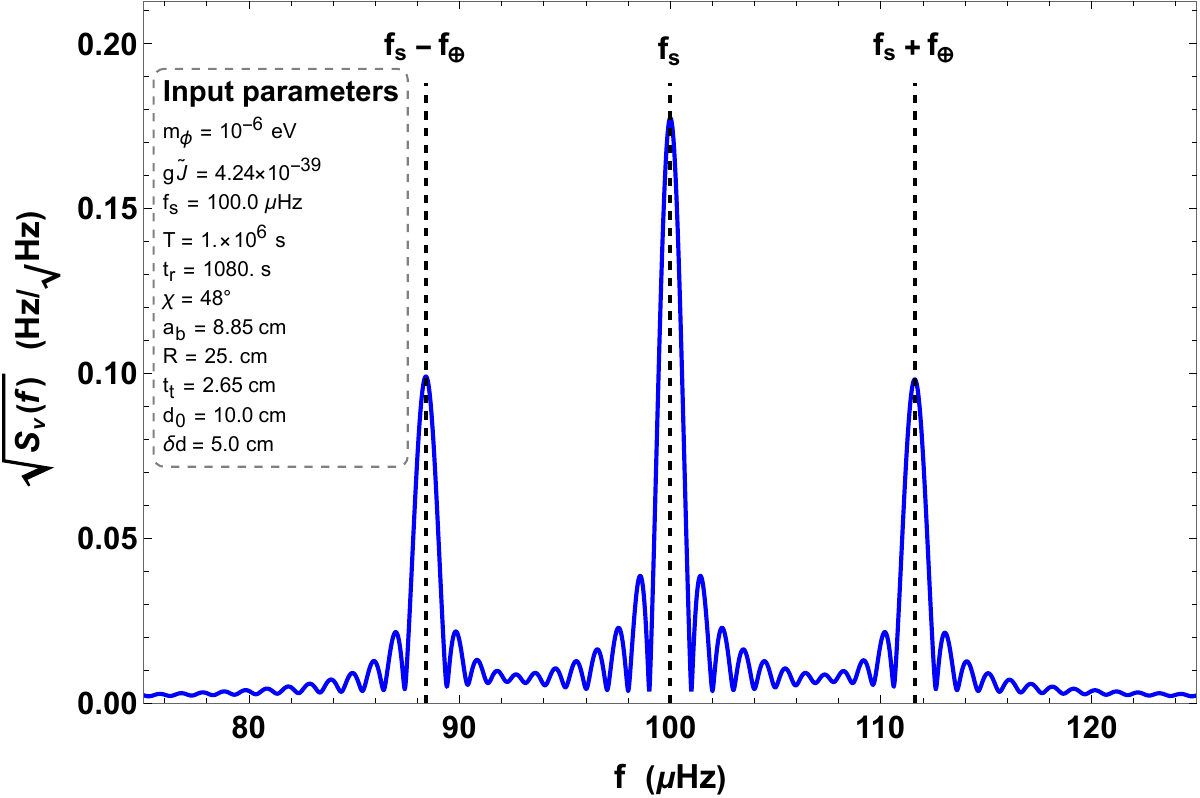}
    \caption{Amplitude spectral density $\sqrt{S_\nu(f)}$ of the scalar-mediated LV contribution to the hydrogen maser Zeeman frequency for a periodically modulated external source. The carrier at $f_s$ and the two sidereal sidebands at $f_s\pm f_\oplus$ provide the characteristic three peak signature. The input parameters are also given in the figure.}
    \label{fig5}
\end{figure}

Figure.~\eqref{fig5} illustrates the characteristic frequency domain signature of the scalar-mediated Lorentz-violating interaction in the presence of a periodically moving source. The source modulation generates a carrier at $f_s$, while the sidereal rotation of the laboratory modulates the orientation of the Lorentz-violating background and produces two sidebands at $f_s\pm f_\oplus$. Thus, rather than a single modulation peak, the signal exhibits a characteristic three peak signature separated by the Earth's modulation frequency. The quantity $\bar{I}_1(m_\phi)$ and hence the overall peak heights, depends on the scalar mass and the source-detector geometry. The peak amplitude spectral density additionally scales as $\sqrt{T}$, whereas the spectral width scales approximately as $1/T$. The oscillatory structures surrounding the three peaks arise from the finite observation window and are not additional physical sidebands.

The spectral signature also distinguishes the scalar-mediated Lorentz-violating interaction from a conventional axion-mediated monopole-dipole interaction \cite{Crescini:2016lwj}. For a periodically moving source with a fixed source-spin orientation, the latter is predominantly source synchronous, producing a signal at $f_s$ and possible higher harmonics $nf_s, (n=2,3,4,\cdots)$. In contrast, the LV background is fixed in the Sun-centered frame, and the Earth's rotation introduces a sidereal modulation. Its mixing with the source modulation produces the characteristic triplet $f_s$ and $f_s\pm f_\oplus$. The fixed sidereal separation of the sidebands therefore provides a characteristic signature of the LV interaction. In FIGs.~\eqref{couplingperp} and \eqref{couplingz}, we present the projected sensitivities to the Lorentz-violating coupling combinations $g_N(\tilde{J}^e_\perp+\tilde{J}^p_\perp)$ and $g_N(\tilde{J}^e_Z+\tilde{J}^p_Z)$ for finite scalar masses, represented by the solid sky-blue and solid yellow curves, respectively, as obtained from hydrogen maser spectroscopy. For the Lorentz-violating coupling components perpendicular to the Earth's rotation axis, we present the joint sensitivity obtained by combining the two sideband contributions, with the coupling normalized by the factor $\sqrt{\eta_+^2+\eta_-^2}$.

\section{Scalar-mediated Lorentz violation with molecular spin precession}\label{molecule-section}
\begin{figure}
    \centering
    \includegraphics[width=12cm]{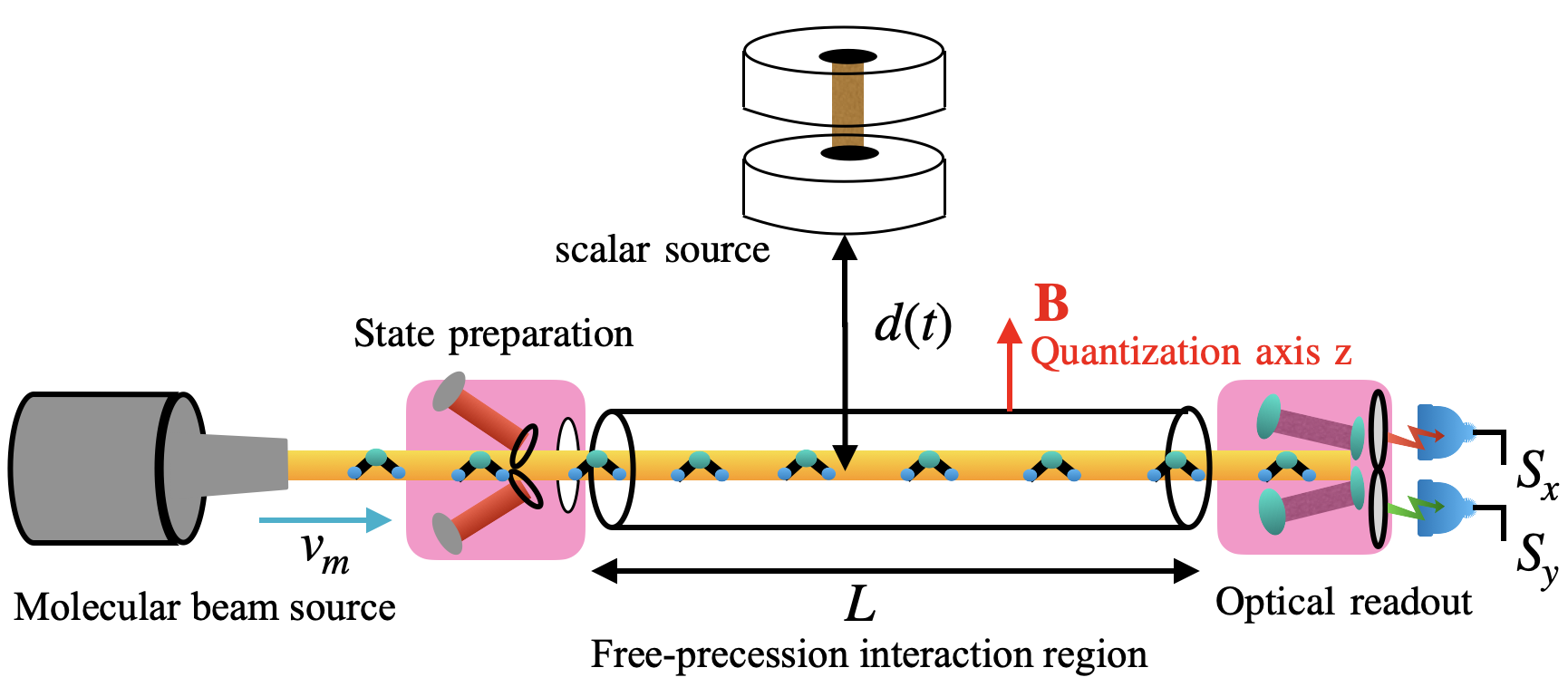}
    \caption{Schematic of the molecular precession setup with a sinusoidally modulated source mass
    }
    \label{figpres1}
\end{figure}
In FIG.~\eqref{figpres1}, we show the schematic diagram of a molecular precession setup with a sinusoidally modulated source which could be used to probe scalar-mediated Lorentz-violating interactions. A molecular beam is prepared in a coherent superposition of two opposite spin states and propagates through a free precession region located near an unpolarized source mass. The source generates a scalar field that induces a LV-dependent energy splitting between the two states, which leads to an accumulated Lorentz-violating relative phase during the molecular flight. The source is sinusoidally moved to modulate the scalar signal and reject any static backgrounds. At the end of the interaction region, the molecular state is analyzed optically and the LV-induced phase is extracted from the fluoroscence asymmetry $\mathcal{A}=\tfrac{S_x-S_y}{S_x+S_y}$, where $S_x$ and $S_y$ denote the fluorescence signals associated with two complementary molecular spin state detection channels following the final Ramsey analysis pulse.

Consider two molecular states $|+\rangle$ and $|-\rangle$ with opposite effective electron spin projections along the experimental quantization axis $\hat{z}$. The corresponding shifts in the energies of these states due to Lorentz-violating interactions are  
\begin{equation}
\delta E_\pm=-g_N I_s(\mathbf{r},m_\phi)\tilde{J^e_z}\langle\pm|\sigma_z|\pm\rangle,
\label{mol1}
\end{equation}
which follows from Eq.~\eqref{eq:a3}, where $I_s$ is given in Eq.~\eqref{eq:a4}. It is useful to define the molecular spin polarization factor 
\begin{equation}
P_e^z=\frac{1}{2}[\langle+|\sigma_e^z|+\rangle-\langle-|\sigma_e^z|-\rangle],
\label{mol2}
\end{equation}
so that the Lorentz-violating energy difference along $\hat{z}$ is 
\begin{equation}
\Delta E_{\mathrm{LV}}(x,t)=\delta E_+-\delta E_-=-2g_N I_s(x,m_\phi)P_e^z \tilde{J^e_z}(t).  
\label{mol3}
\end{equation}

Note, in the ideal fully polarized case $P_e^z=1$ as $\sigma_e^z|+\rangle=+|+\rangle$ and $\sigma_e^z|-\rangle=-|-\rangle$. However, for a real molecular state, $P_e^z$ need not be one. For the stretched $M_J=\pm 1$ states of the ThO $H\,{}^3\Delta_1$, $J=1$ manifold, the relevant electron spin expectation value changes sign between the two states and is close to its maximal value under strong molecular projection. These states are therefore particularly sensitive to electron spin dependent interactions. We retain the factor $P_e$ explicitly to account for the detailed molecular polarization and to keep the formalism applicable to all other molecular species.

The experiment initially prepares a coherent superposition
\begin{equation}
|\psi(0)\rangle=\frac{1}{\sqrt{2}}(|+\rangle+|-\rangle).   
\label{mol4}
\end{equation}
After free evolution for a time $\tau$,
\begin{equation}
|\psi(\tau)\rangle=\frac{1}{\sqrt{2}}[e^{-iE_+\tau}|+\rangle+e^{-iE_-\tau}|-\rangle].    
\label{mol5}
\end{equation}

For molecules propagating through the interaction region with velocity $v_m$, the Lorentz-violating energy splitting generates a relative phase between the two spin states. The accumulated phase is
\begin{equation}
\Phi_{\mathrm{LV}}=\int \Delta E_{\mathrm{LV}}(t)dt=-2g_NP_e^z\int ^L_0 I_s(x,m_\phi) \tilde{J_z^e } (t_x)\frac{dx}{v_m},
\label{mol6}
\end{equation}
where $t_x=t_0+\tfrac{x}{v_m}$ is the time at which a given molecule reaches a position $x$ along the beam line and $t_0$ is the time when the molecule enters the precession region, usually taken at $x=0$. Since, the molecular transit time is much shorter than the characteristic variation time of the laboratory frame LV coefficient, $\tilde{J^e_z}$ may be regarded as constant during a single transit. Introducing the path-averaged source profile
\begin{equation}
\bar{I}_s(m_\phi) =\frac{1}{L}\int^L_0  I_s(m_\phi)dx,
\label{mol7}
\end{equation}
we obtain the corresponding anomalous angular frequency as
\begin{equation}
\Phi_\mathrm{LV}=-2g_N  \bar{I}_s(m_\phi)P_e^z \tilde{J^e_z}\frac{L}{v_m}=\delta\omega_{\mathrm{LV}}\tau_\mathrm{coh},
\label{mol8}
\end{equation}
where $\tau_{\mathrm{coh}}\simeq L/v_m$ is called the spin coherence (precession) time. At the end of the precession region, the molecular state is projected onto spin-quadrature $x$ and $y$ states and read out optically. The experiment uses a readout laser to perform this projection. The number of fluorescence photons detected when the readout is configured for the $X$ analyzer gives $S_x$ and the fluorescence obtained with the orthogonal $Y$ analyzer gives $S_y$.

Suppose the total accumulated spin phase is $\vartheta=\vartheta_0+\Phi_{\mathrm{LV}}$, where $\vartheta_0$ denotes the conventional phase arising from the experimental setup. After free precession, the molecule is projected onto two orthogonal analysis states, which we can denote $X$ and $Y$. Ideally, their detection probabilities are complementary, given as
\begin{equation}
P_X=\frac{1}{2}(1+C\cos\vartheta),~~~P_Y=\frac{1}{2}(1-C\cos\vartheta),  
\label{mol9}
\end{equation}
where $0\leq C\leq 1$ is the fringe contrast or visibility. The maximum value $C=1$ corresponds to an ideal experiment with perfect state preparation, coherence, projection, and detection. In reality, imperfect preparation/readout, decoherence, finite laser efficiency, etc., can reduce it.

Consider $S_0$ denote the total fluorescence signal that would be obtained from the molecular ensemble in that measurement cycle. Therefore, we can write
\begin{equation}
S_x=\frac{S_0}{2} [1+C\cos(\vartheta_0+\Phi_{\mathrm{LV}})],~~~ S_y=\frac{S_0}{2} [1-C\cos(\vartheta_0+\Phi_{\mathrm{LV}})],  
\label{mol10}
\end{equation}
where $S_0$ is simply the overall signal normalization. In the idealized scenario $S_0=S_x+S_y$, however realistically $S_0$ depends on the number of molecules passing through the apparatus, optical excitation efficiency, fluorescence collection efficiency, and detector efficiency. 

We define the normalized difference
\begin{equation}
\mathcal{A}=\frac{S_x-S_y}{S_x+S_y}=C\cos(\vartheta_0+\Phi_{\mathrm{LV}})\approx C\cos\vartheta_0-C\sin\vartheta_0\Phi_{\mathrm{LV}}, 
\label{mol11}
\end{equation}
where the approximation in Eq.~\eqref{mol11} holds for $|\Phi_{\rm{LV}}|\ll1$. Thus, the change in $\mathcal{A}$ is governed by the Lorentz-violating interaction $\delta \mathcal{A}=-C\sin\vartheta_0\Phi_{\rm{LV}}$. The experiment chooses $\vartheta_0$ near the steepest part of the fringe so that a tiny phase produces the largest possible change in fluorescence asymmetry. The contrast $C$ can be measured from the Ramsey fringe as
\begin{equation}
C=\frac{S_{\rm{max}}-S_{\rm{min}}}{S_{\rm{max}}+S_{\rm{min}}}, 
\label{mol12}
\end{equation}
where $S_{\rm{max}}(S_{\rm{min}})$ are obtained by taking the maximum (minimum) value of $S_x$, given in Eq.~\eqref{mol10}. Operating at the point of maximal slope of $|\sin\vartheta_0|\simeq1$ gives
\begin{equation}
|\delta \mathcal{A}|\simeq C|\Phi_{\mathrm{LV}}|=C|\delta\omega_{\mathrm{LV}}|\tau_\mathrm{coh}.    
\label{mol13}
\end{equation}
Thus, the experiment does not have to observe a spin vector making a complete rotation. It measures the small relative quantum phase accumulated between the two opposite spin components during the molecular flight.

\begin{figure}
    \centering
    \includegraphics[width=12cm]{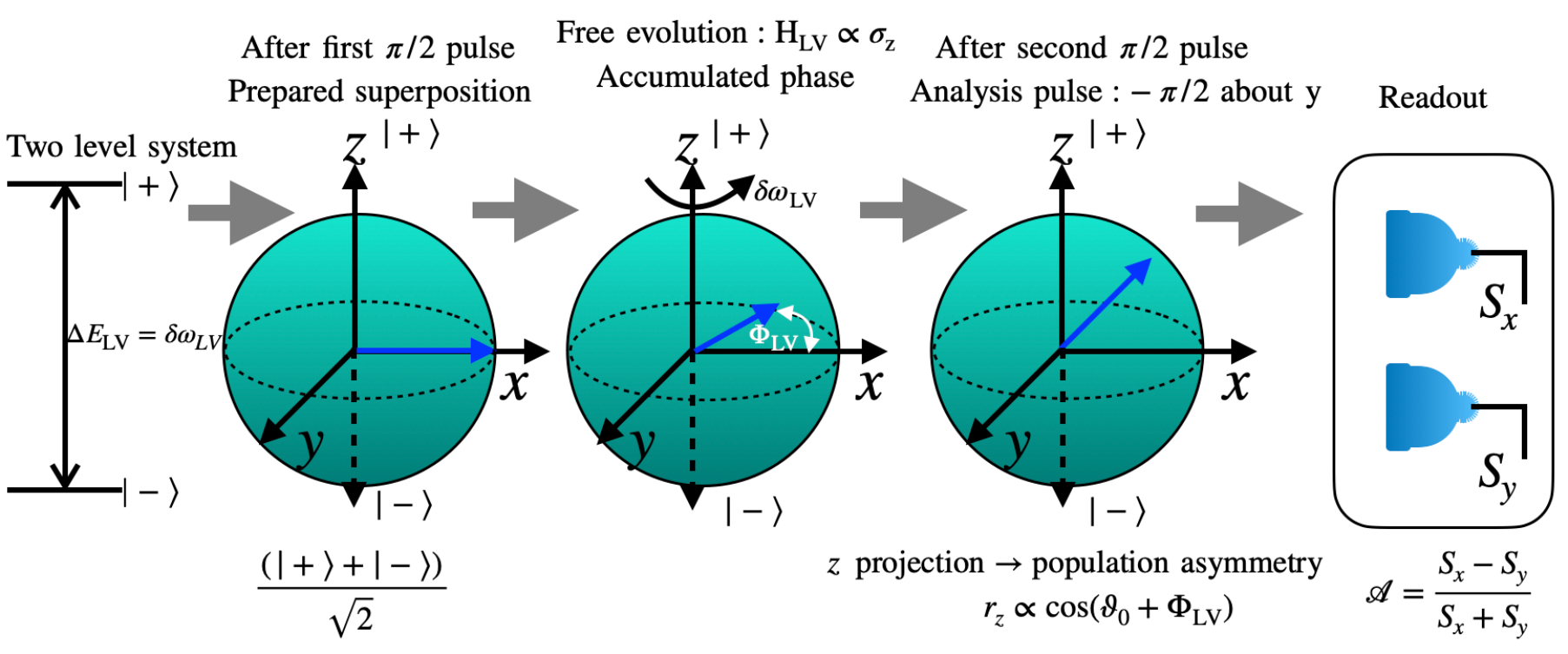}
    \caption{Bloch sphere representation for Ramsey spin precession probing Lorentz-violating interaction}
    \label{figpres2}
\end{figure}

Figure~\eqref{figpres2} shows the Bloch sphere (geometrical illustration of the quantum state of a two level system) representation for Ramsey spin precession probing Lorentz-violating interaction. The molecule is initially prepared in the basis state $|\psi_{\rm{int}}\rangle=|+\rangle$ which is along $+\hat{z}$ on the Bloch sphere, i.e., pointing towards the north pole given as $\mathbf{r}=(0,0,1)$. The first $\pi/2$ pulse prepares the molecule in the coherent superposition 
\begin{equation}
|\psi_0\rangle=\frac{|+\rangle+|-\rangle}{\sqrt{2}},
\label{mol14}
\end{equation}
corresponding to a Bloch vector along $\hat{x}$, i.e., $\mathbf{r}=(1,0,0)$. During free evolution, the Lorentz-violating interaction $H_{\rm{LV}}\propto \sigma_z$ produces an energy splitting $\Delta E_{\mathrm{LV}}=\delta\omega_{\mathrm{LV}}$. Therefore, the two states acquire a relative phase $\Phi_{\rm{LV}}=\delta\omega_{\rm{LV}}\tau_{\rm{coh}}$. Thus, the Lorentz-violating interaction produces only an azimuthal rotation in the $x-y$ plane during free evolution. The corresponding Bloch vector including a controllable reference phase $\vartheta_0$ takes the form
\begin{equation}
\mathbf{r}=[\cos(\vartheta_0+\Phi_{\rm{LV}}),\sin(\vartheta_0+\Phi_{\mathrm{LV}}),0].   
\label{mol15}
\end{equation}
The second $-\pi/2$ pulse is applied about $y$, i.e., $R_y(-\pi/2):(x,y,z)\to (-z,y,x)$. Since the Bloch vector initially lies in the equatorial plane $z=0$, before the pulse, its position after the second pulse is given by
\begin{equation}
\mathbf{r}_f=[0,\sin(\vartheta_0+\Phi_{\mathrm{LV}}),\cos(\vartheta_0+\Phi_{\rm{LV}})].
\label{mol16}
\end{equation}
Therefore, $r_z=\cos(\vartheta_0+\Phi_{\rm{LV}})$, which is the population imbalance after applying the analysis pulse i.e., $r_z=P_+-P_-$, where $P_\pm=(1\pm r_z)/2=[1\pm\cos(\vartheta_0+\Phi_\mathrm{LV})]/2$. The two $\pi/2$ pulses are created by Rabi oscillation driven by a resonant electromagnetic field.

Thus, the optical readout measures these populations through the two fluorescence channels, giving schematically 
\begin{equation}
\mathcal{A}=\frac{S_x-S_y}{S_x+S_y}\simeq Cr_z=C\cos(\vartheta_0+\Phi_{\rm{LV}}).   
\label{mol17}
\end{equation}
where $S_{x(y)}=\eta_{x(y)}NP_{+(-)}$, with $N$ is the number of detected molecules and $\eta_{x(y)}$ contain the excitation, collection and detector efficiency. The second $-\pi/2$ pulse converts the tiny LV-induced relative phase into a measurable population or fluorescence asymmetry. We assume $\eta_x=\eta_y$ if the two channels are balanced and Eq.~\eqref{mol17} is the asymmetry formula which we used in Eq.~\eqref{mol11}. 

\subsection{Sensitivity to the Lorentz-violating coupling from molecular beam}\label{molecular-beam}
Following the approach adopted in hydrogen maser spectroscopy, we consider a sinusoidally oscillating external source in the molecular Ramsey spectroscopy setup to assess its sensitivity to scalar-mediated Lorentz-violating interactions. We allow the source mass to oscillate as 
\begin{equation}
d(t)=d_0+\delta d\cos(\omega_s t)   
\label{mol18}
\end{equation}
where $d_0$ is the mean distance between the source surface and the molecular beam axis. 
For the experimentally relevant regime $f_s\tau_{\rm{coh}}\ll 1$, the source position changes negligibly during a single Ramsey interrogation. Consequently, the scalar field is effectively static for each individual molecule and $\Phi_{\rm{LV}}(t_0)\simeq \delta\omega_{\rm{LV}}(t_0)\tau_{\rm{coh}}$. Successive molecular packets nevertheless sample different phases of the slowly varying source motion, allowing the modulation to be reconstructed over the full data taking period.

In a molecular beam experiment, the molecules propagate over a finite precession length $L$, and therefore sample a spatially varying scalar field along their trajectory. For a narrow molecular beam propagating along the $x$ direction with velocity $v_m$, we define the effective source integral sampled during the Ramsey free precession interval as 
\begin{equation}
\overline{I}_s(d,m_\phi)=\frac{1}{L}\int^{L/2}_{-L/2}dx I_s (x,d,m_\phi) 
\label{mol19}
\end{equation}
where the expression of $I_s$ is given in Eq.~\eqref{eq:a4}. If the finite transverse beam profile is also be included in the source integral, then Eq.~\eqref{mol19} becomes
\begin{equation}
\overline{I}_s(d,m_\phi)=\frac{1}{L}\int^{L/2}_{-L/2}dx\int d^2\rho W(\bm{\rho})I_s(x,\bm{\rho}, d,m_\phi), 
\label{mol20}
\end{equation}
where $W(\bm{\rho})$ is normalized according to $\int d^2\rho W(\bm{\rho})=1$. For a sufficiently narrow beam, the transverse average maybe neglected. For a cylindrical source of radius $R$, thickness $t_t$, and uniform nucleon density $n_N$, the source integral below the cylinder axis is 
\begin{equation}
I_s(d,m_\phi)=\frac{n_N}{2m_\phi}\int ^{d+t_t}_d dz[e^{-m_\phi z}-e^{-m_\phi\sqrt{z^2+R^2}}].  
\label{mol21}
\end{equation}
Away from the symmetry axis, as required for the longitudinal path average, the full finite-cylinder integral is given as
\begin{equation}
I_s(x,d,m_\phi)=\frac{n_N}{4\pi}\int ^{d+t_t}_d dz^\prime \int ^R_0 \rho^\prime d\rho^\prime \int ^{2\pi}_0 d\varphi^\prime \frac{e^{-m_\phi \mathcal{R}}}{\mathcal{R}}, 
\label{mol22}
\end{equation}
where $\mathcal{R}=\sqrt{(x-\rho^\prime\cos\varphi^\prime)^2+{\rho^\prime}^2{\sin^2}\varphi^\prime+{z^\prime}^2}$ and Eq.~\eqref{mol22} needs to be calculated numerically. Since, the source is moving sinusoidally, the trajectory averaged source integral is periodic and maybe expanded as 
\begin{equation}
\overline{I}_s[d(t)]=\bar{I}_0 +\bar{I}_1\cos\omega_s t+ \bar{I}_2\cos2\omega_s t+\cdots,  
\label{mol23}
\end{equation}
where
\begin{equation}
\bar{I}_1(m_\phi)=\frac{1}{\pi}\int ^{2\pi}_0 d\theta \overline{I_s}(d_0+\delta d\cos\theta,m_\phi)\cos\theta,~~\theta=\omega_s t.  
\label{mol24}
\end{equation}
Combining the source motion and trajectory averages, Eq.~\eqref{mol24} becomes 
\begin{equation}
\bar{I}_1(m_\phi)=\frac{1}{\pi L}\int^{L/2}_{-L/2} dx\int ^{2\pi}_0 d\theta I_s[x, d_0+\delta d\cos\theta,m_\phi]\cos\theta.
\label{mol25}
\end{equation}
Therefore, Eq.~\eqref{mol3} becomes
\begin{equation}
\Delta E_{\rm{LV}}(t)=\delta\omega_{\rm{LV}}(t)=-2g_N P_e^z \tilde{J}_z^e(t) \overline{I}_s[d(t)]\simeq -2g_N P_e^z \tilde{J}_z^e(t) \overline{I}_1\cos\omega_s t,
\label{mol26}
\end{equation}
where the approximation denotes that we only keep the relevant source-correlated first harmonic. Using Eq.~\eqref{eq:16}, we can write Eq.~\eqref{mol26} in terms of the Lorentz-violating coefficients in the Sun-centered frame as
\begin{align}
\delta\omega_{\rm{LV}}(t)&=-2g_N P_e^z \overline{I}_1J_Z^e\cos\chi \cos\omega_s t-g_N P_e^z\overline{I}_1 J^e_X\sin\chi[\cos\omega_+t+\cos\omega_- t]\nonumber\\
&-g_N P_e^z\overline{I}_1 J^e_Y\sin\chi[\sin\omega_+t-\sin\omega_- t],
\end{align}
with $\omega_\pm=\omega_s\pm \omega_\oplus$. This is the similar sideband structure as we already obtain in Eq.~\eqref{eq:a13} for the hydrogen maser spectroscopy scenario. Thus, experimentally achievable sensitivity on the fluorescence asymmetry can directly constrain Lorentz-violating coupling from Eq.~\eqref{mol13}.

We adopt the same source modulation parameters as in the hydrogen Zeeman spectroscopy analysis, namely $f_s=100~\mu{\rm Hz},~
t_r=18~{\rm min}$, for which $\eta_s\simeq 0.981$, giving $f_+=111.6~\mathrm{\mu Hz}$, $f_-=88.4~\mathrm{\mu Hz}$, with $\eta_-\simeq 0.985$, $\eta_+\simeq 0.976$. The corresponding sensitivities of the Lorentz-violating coupling becomes
\begin{equation}
|g_N\tilde{J}^e_Z|\sim \frac{\sigma_\omega}{2P_e^z|\bar{I}_1||\cos\chi|\eta_s},~~~|g_N\tilde{J}_\perp^e|_\pm\sim \frac{\sigma_\omega}{P_e^z|\bar{I}_1||\sin\chi|\eta_\pm},    
\end{equation}
where $\sigma_\omega$ is the sensitivity to the anomalous angular precession frequency. It is related to the Ramsey asymmetry sensitivity by
\begin{equation}
\sigma_{\mathcal A}
=C\tau_{\rm coh}|\sin\vartheta_0|\,\sigma_\omega .
\end{equation}
For the molecular beam we take $\tau_{\rm{coh}}=5~\mathrm{ms}$, which is much shorter than $t_r$. Since $\pi f_s\tau_\mathrm{coh}\simeq 1.6\times 10^{-6}$, the transit time attenuation is negligible, given as $\eta_\mathrm{transit}=\tfrac{\sin (\pi f_s\tau_\mathrm{coh})}{\pi f_s\tau_\mathrm{coh}}\simeq 1$. The source advances by only $\Delta\phi_s=2\pi f_s\tau_{\mathrm{coh}}\simeq 3.1\times 10^{-6}~\mathrm{rad}$, during one molecular interrogation. For $\delta d=5~\mathrm{cm}$, the maximum source velocity is $v_\mathrm{max}=31~\mathrm{\mu m/s}$ corresponding to a displacement of only $0.16~\mathrm{\mu m}$ during $5~\mathrm{ms}$. Thus, the scalar field is effectively static for each molecular packet, while successive packets sample the slow source modulation. We also take $T\simeq  11.6~\mathrm{days}$, giving a Fourier resolution $\Delta f=1~\mathrm{\mu Hz}$, sufficient to resolve the carrier and sidereal sidebands.

We choose the benchmark parameters of the cylindrical tungsten source as $M_s=100~\mathrm{kg}$, $R=25~\mathrm{cm}$, $t_t=2.65~\mathrm{cm}$, $d_0=10~\mathrm{cm}$, $\delta d=5~\mathrm{cm}$, $L=1~\mathrm{m}$, which leads to $|\bar{I}_1(m_\phi\to 0)|\simeq 4.64\times 10^{11}~\mathrm{GeV}$. Using ACME-III benchmark $\delta\omega=15~\mathrm{\mu rad~s^{-1}}$ \cite{Agrawal:2023lmw}, and $P_e^z\simeq 1$, we obtain the sensitivities on the Lorentz-violating coupling as
\begin{equation}
|g_N\tilde{J}_Z^e|\sim 1.6\times 10^{-41},~~~|g_N\tilde{J}_\perp^e|_-\sim 2.91\times 10^{-41}, ~~~|g_N\tilde{J}_\perp^e|_+\sim 2.94\times 10^{-41},  
\end{equation}
which is valid for the scalar mass $m_\phi\ll 1/\mathcal{R}_\mathrm{max}\sim 2.6\times 10^{-7}~\mathrm{eV}$, since, $\mathcal{R}_\mathrm{max}=\sqrt{[(L/2)+R]^2+(d_{\mathrm{max}}+t_t)^2}$. For a nonzero scalar mass, the source integral must be evaluated numerically. In the regime $m_\phi\gg 1/\mathcal{R}_{\mathrm{max}}$, the finite interaction range leads to Yukawa suppression, thereby weakening the sensitivity to the scalar coupling. In FIGs.~\eqref{couplingperp} and \eqref{couplingz}, we present the projected sensitivities to the Lorentz-violating couplings $|g_N\tilde{J}_\perp^e|$ and $|g_N\tilde{J}_Z^e|$ for finite scalar masses, represented by the solid green and solid orange curves, respectively, as obtained from spin-precession measurements using ThO molecules. We present the combined sensitivity obtained from both sideband contributions, incorporating the normalization factor $\sqrt{\eta_+^2+\eta_-^2}$ in the coupling sensitivity.

\subsection{Sensitivity on the Lorentz violating coupling from trapped molecules}\label{ultracold-section}

The sensitivity to Lorentz-violating couplings can be further improved by employing ultracold atoms or molecules \cite{Carr:2009iuf,Jenkins:2026rko}, for which coherence times of order $10-100~\mathrm{s}$ may be achievable. Since the accumulated Lorentz-violating phase, $\Phi_{\mathrm{LV}}$, scales linearly with the coherent interrogation time, longer coherence times offer improved sensitivity to LV interactions. Moreover, trapping allows ultracold molecules to remain in close proximity to the source, overcoming the transit time limitations of molecular beam experiments \cite{PhysRevResearch.3.033013}. Unlike the free-precession molecular beam configuration, where the source-induced interaction must be averaged along the molecular trajectory (cf. Eq.~\eqref{mol19}), a trapped molecular ensemble experiences an approximately fixed source-sensor separation during each individual Ramsey interrogation. This eliminates longitudinal trajectory averaging and enables the molecules to accumulate the Lorentz-violating phase over substantially longer interrogation times, potentially providing a significant improvement in sensitivity. In such Ramsey experiments, the trapping laser can be operated near a magic wavelength, where the two Ramsey states experience nearly identical polarizabilities. This suppresses differential AC Stark shifts and therefore minimizes spurious phase accumulation during the interrogation.

For the same $100~\mathrm{kg}$ cylindrical tungsten source of $R=25~\mathrm{cm}$, $t_t\simeq 2.65~\mathrm{cm}$, with sinusoidal motion $d(t)=d_0+\delta d\cos\omega_s t$, with $d_0=10~\mathrm{cm}$ and $\delta d=5~\mathrm{cm}$, we can treat a sufficiently compact ultracold cloud as a pointlike detector located on the cylinder symmetry axis. Consider the molecule is located at $\mathbf{r}=(x,y,z)$ and hence the scalar field generated by the source can be written as
\begin{equation}
\phi(\mathbf{r},t)=g_N I_s(\mathbf{r},d(t),m_\phi),
\label{ultra1}
\end{equation}
where the expression of $I_s(\bm{r},d(t),m_\phi)$ is given in Eq.~\eqref{mol22} with the general expression $\mathcal{R}=\sqrt{(x-\rho^\prime\cos\varphi^\prime)^2+(y-\rho^\prime\sin\varphi^\prime)^2+(z-z^\prime)^2}$. If the molecule is at the center of the trap i.e., $\mathbf{r}=\mathbf{0}$, the cylindrical symmetry allows the transverse integrals to be performed analytically. Therefore, the solution of Eq.~\eqref{mol22} becomes Eq.~\eqref{mol21}, which gives the instantaneous scalar source integral seen by a molecule localized at the center of the trap. For a finite ultracold molecular cloud, the total number of molecules for a given molecular density $n_\mathrm{mol}(\mathbf{r})$ is obtained as
\begin{equation}
N_\mathrm{mol}=\int d^3 r\,n_{\mathrm{mol}}(\mathbf{r}), 
\label{ultra2}
\end{equation}
and the source integral averaged over the trapped ensemble becomes
\begin{equation}
\overline{I}_s^{\mathrm{trap}}(d(t),m_\phi)=\frac{1}{N_\mathrm{mol}} \int d^3 r \,  n_{\mathrm{mol}}(\mathbf{r}) I_s(\mathbf{r},d(t),m_\phi).
\label{ultra3}
\end{equation}
For a thermal molecular ensemble confined in an approximately harmonic trap, a typical density profile can be written as \cite{Grimm:1999hsz}
\begin{equation}
n_\mathrm{mol}(\mathbf{r})=\frac{N_{\rm{mol}}}{(2\pi)^{3/2}\sigma_x\sigma_y\sigma_z}\exp\Big[-\frac{x^2}{2\sigma_x^2}-\frac{y^2}{2\sigma_y^2}-\frac{z^2}{2\sigma_z^2}\Big],    
\label{ultra4}
\end{equation}
where 
\begin{equation}
\sigma_i=\sqrt{\frac{T}{m_\mathrm{mol}\omega^2_i}},~~~i=x,y,z,
\label{ultra5}
\end{equation}
for a dilute thermal cloud in a harmonic potential, with $T$ denotes the temperature of the trapped ultracold molecular ensemble, $m_\mathrm{mol}$ denotes the mass of one molecule and $\omega_i$ are the harmonic trapping frequencies along the three spatial directions. 

The sinusoidal motion of the macroscopic source periodically modulates the scalar field and hence Eq.~\eqref{ultra3} varies as Eq.~\eqref{mol23}, where the amplitude of the first harmonic is given as Eq.~\eqref{mol24}. Therefore, the molecular cloud average explicitly gives the amplitude of the first harmonic as 
\begin{equation}
I_1^{\mathrm{trap}}(m_\phi)=\frac{1}{\pi N_{\mathrm{mol}}}\int ^{2\pi}_0 d\theta\cos\theta\int d^3 r\, n_{\mathrm{mol}}(\mathbf{r})I_s(\mathbf{r},d_0+\delta d\cos\theta,m_\phi).
\label{ultra6}
\end{equation}
However, in practice, the ultracold cloud can be much smaller than the characteristic source-sensor separation. In fact, for a moderately tight optical trap $\mu \mathrm{K}$ molecular cloud with $f_i=100~\mathrm{Hz}$, we have $\sigma_i\simeq 10~\mathrm{\mu m}$. Therefore, if $\sigma_{x,y,z}\ll d_{\mathrm{min}}$, $d_{\mathrm{min}}=d_0-\delta d$, the scalar field varies negligibly across the cloud. 

Expanding about the trap center, we obtain the source integral as 
\begin{equation}
I_s(\mathbf{r},d)=I_s(0,d)+r_i\partial_i I_s(0,d)+\frac{1}{2}r_ir_j\partial_i\partial_jI_s(0,d)+\cdots. 
\label{ultra7}
\end{equation}
For a symmetric cloud $\langle r_i\rangle=0$, and the leading correction of Eq.~\eqref{ultra7} is second order in the cloud size. Consequently, 
\begin{equation}
\overline{I}_1^\mathrm{trap}(m_\phi)\simeq I_1^\mathrm{trap}(m_\phi)= \frac{1}{\pi}\int ^{2\pi}_0 d\theta I_s(0,d_0+\delta d\cos\theta,m_\phi)\cos\theta.
\label{ultra8}
\end{equation}
Thus, unlike the molecular beam setup, there is no longitudinal trajectory average. The molecules remain localized near the source throughout the interrogation. Therefore, in the long range limit $m_\phi\to 0$, we obtain the value for the source integral calculated numerically as $|\overline{I}_1^{\mathrm{trap}}(0)|\simeq 8.9\times 10^{11}~\mathrm{GeV}$, which is about twice the trajectory-averaged molecular beam value because the ultracold molecules remain localized close to the source rather than spending much of the interrogation far from it.

For the longitudinal Lorentz-violating coefficient in the Sun-centered frame, the carrier at $f_s$ has amplitude
\begin{equation}
\delta\omega_Z=2g_N P_e^z|\overline{I}_1^\mathrm{trap}| |\tilde{J}_Z^e|  |\cos\chi|\eta_s. 
\label{ultra9}
\end{equation}
Similarly, for the transverse coefficients, the modulation is split into the two sidebands with frequencies $f_s\pm f_\oplus$ and the amplitude of each sideband is 
\begin{equation}
\delta\omega_\perp= g_N P_e^z|\overline{I}_1^\mathrm{trap}| |\tilde{J}_\perp^e|  |\sin\chi|\eta_\pm.   
\label{ultra10}
\end{equation}
Using the same input parameters considered in the previous section, $P_e^z\simeq 1$, $\chi=48^\circ$, $\eta_s=0.981$, $\eta_+=0.976$, $\eta_-=0.985$, and equating Eqs.~\eqref{ultra9} and \eqref{ultra10} with the optimistic ultracold molecule sensitivity $\delta\omega=1~\mathrm{nrad/s}$ \cite{Agrawal:2023lmw}, we obtain

\begin{equation}
|g_N\tilde{J}_Z^e|\sim 5.65\times 10^{-46},~~~|g_N\tilde{J}_\perp^e|_-\sim 1.01\times 10^{-45},~~~|g_N\tilde{J}_\perp^e|_+\sim 1.02\times 10^{-45}. 
\label{ultra11}
\end{equation}
The limits are valid for the mass of the scalars $m_\phi\ll 1/\mathcal{R}_\mathrm{max}\sim 6.45\times 10^{-7}~\mathrm{eV}$, since $\mathcal{R}_\mathrm{max}=\sqrt{R^2+(d_\mathrm{max}+t_t)^2}$.

Typically, the shot noise limited frequency sensitivity scales as $\delta\omega\sim \tfrac{1}{\tau_\mathrm{coh} \sqrt{N}}$, where $N$ is the total number of detected ultracold molecules contributing independent measurements. In practice, if $N_0$ denotes the number of molecules prepared per experimental cycle, $T_{\mathrm{int}}$ is the total integration time and $\eta$ denotes the detection efficiency, then the shot noise uncertainty becomes $\delta\omega\simeq \frac{1}{C\sqrt{\eta \tau_{\mathrm{coh}}T_{\mathrm{int}}N_0}}$, where $C\leq 1$ denotes the Ramsey contrast. Using $C=1$, $\eta=0.1$, $T_{\mathrm{int}}=10^7~\mathrm{s}$, $\tau_{\mathrm{coh}}=100~\mathrm{s}$, the spin-precession frequency can be in the range $\delta\omega\in (100-1)~\mathrm{nrad/s}$ for $N_0\in(10^6-10^{10})$.

Thus, increasing $N_0$, together with a long coherence time, can substantially improve the sensitivity to small spin precession frequencies, potentially from the $\mathcal{O}(10)~\mu\mathrm{rad/s}$ regime to the $\mathcal{O}(100-1)~n\mathrm{rad/s}$ level. Realizing this improvement requires high preparation and detection efficiencies while suppressing collisional effects, density dependent decoherence, and technical noise.

For our sensitivity projections, we adopt an optimistic benchmark of $\delta\omega=1\,\mathrm{nrad/s}$, motivated by the ultracold molecular spin-precession proposal of Ref.~\cite{Agrawal:2023lmw}. Achieving this sensitivity would require long coherence times ($\tau_{\rm coh}\sim100\,\mathrm{s}$), large molecular ensembles, and efficient state preparation and readout, together with stringent control of magnetic and other systematic backgrounds. In particular, the molecular states must have appreciable electron spin polarization and suitable co-magnetometry must be implemented to distinguish the signal from magnetic field fluctuations. While such sensitivity remains experimentally ambitious and has not yet been demonstrated, it provides a useful benchmark for the potential reach of future ultracold molecular experiments.

Since the source is modulated at $f_s$, the scalar field varies slightly during each Ramsey interrogation. The corresponding finite interrogation time attenuation is $\eta_\mathrm{coh}(f)=\tfrac{\sin(\pi f\tau_{\mathrm{coh}})}{\pi f\tau_\mathrm{coh}}$. For $f=f_s=100~\mathrm{\mu Hz}$ and $\tau_\mathrm{coh}=10-100~\mathrm{s}$, we obtain $\eta_\mathrm{coh}\simeq 1$, indicating that the source-induced signal is essentially unaffected by averaging over a single interrogation. In FIGs.~\eqref{couplingperp} and \eqref{couplingz}, we present the projected sensitivities to the Lorentz-violating couplings $|g_N\tilde{J}^e_\perp|$ and $|g_N\tilde{J}^e_Z|$ represented by the solid blue and red curves for joint sidebands, respectively using spin precession of the ultracold molecules. These curves demonstrate the strongest sensitivity reach compared with those obtained from hydrogen maser spectroscopy and spin-precession measurements using ThO molecules.

\section{Conclusions and discussions}\label{sec4}

In this work, we investigate the sensitivity of hydrogen maser spectroscopy and spin-precession measurements using atoms and molecules to scalar mediated, spin-dependent Lorentz-violating interactions. We demonstrate that hydrogen maser spectroscopy can probe previously unexplored parameter space for scalar masses of $\mathcal{O}(10~\mathrm{keV})$. Furthermore, introducing a movable external source of scalar fields near the hydrogen maser apparatus extends its sensitivity to ultralight scalars with masses of sub-$\mu\mathrm{eV}$. We also show that spin-precession measurements using atoms and molecules, particularly ultracold molecules with extended coherence times, can significantly improve the sensitivity to these Lorentz-violating couplings, offering a promising avenue for probing previously unconstrained regions of parameter space.

Previous constraints on such mixed Lorentz-violating couplings have primarily been obtained by considering astrophysical bodies as sources of the scalar field. Owing to the large source-detector separations, these constraints are predominantly sensitive to ultralight scalars with masses $m_\phi\lesssim 10^{-10}~\mathrm{eV}$. Limits on individual couplings are also obtained from Red giants for on-shell Lorentz-violating scalar bosons. In contrast, tabletop experiments, with their much shorter source-detector distances, offer the possibility of probing mixed Lorentz-violating couplings for heavier scalar mediators, thereby providing a complementary approach to existing astrophysical constraints. Moreover, the controlled laboratory environment allows greater flexibility in experimental design, precise modulation of the source configuration, and systematic characterization and suppression of background noise, potentially improving the sensitivity to these interactions.

The proposed experimental searches require careful control of systematic backgrounds that may mimic or obscure the Lorentz-violating signal. In hydrogen maser spectroscopy, the dominant sources of systematic uncertainty include magnetic field fluctuations, solenoid current drift, temperature variations, and atomic density dependent frequency shifts \cite{Humphrey:2001wm}, which can be mitigated through magnetic shielding, active field stabilization, thermal control, and continuous monitoring of experimental parameters. The co-magnetometry techniques is useful to suppress the stray magnetic field effects in measuring the spin-precession frequency \cite{Agrawal:2023lmw}. For ThO molecular spin precession measurements, additional backgrounds arise from electric field fluctuations, differential Stark shifts, laser polarization imperfections, and variations in molecular beam properties \cite{ACME:2016sci}, which can be suppressed through stabilized electromagnetic fields, improved optical control, and differential measurements. In ultracold molecular setups, differential trapping-induced Stark shifts, intermolecular interactions, and trap vibrations may limit the coherence time and sensitivity \cite{Anderegg:2023dgq,Gregory:2021bsk,Gregory:2023fxs}, motivating magic trapping conditions, controlled molecular densities, and mechanical isolation. 

An important systematic background common to all three configurations arises from the sinusoidal motion of the external source. In particular, the ARIADNE collaboration investigated magnetic backgrounds associated with a rotating tungsten source mass and demonstrated that residual magnetic impurities can produce source-correlated magnetic fields that may mimic the expected exotic spin-dependent interaction \cite{ARIADNE:2017tdd,ARIADNE:2020cpo}. This motivates the use of high purity tungsten, careful magnetic characterization, demagnetization, and appropriate shielding in the proposed setups. Furthermore, several null tests can be performed, including measurements with the source stationary or removed while operating the translation mechanism, as well as reference measurements using suitably chosen spin states or detection channels. Varying the source-detector separation provides an additional consistency check by testing whether the measured signal follows the predicted scalar-mediated Yukawa dependence. Reversal of the applied magnetic field or molecular spin state preparation can further help identify systematic effects, provided the expected transformation of the Lorentz-violating signal is taken into account. Finally, the characteristic sidereal sidebands at $f_s\pm f_\oplus$, together with the source-modulation frequency $f_s$, offer complementary signatures for distinguishing Lorentz-violating interactions from conventional backgrounds, although source-correlated daily variations must also be controlled.

The proposed experimental setups can also be extended to probe other Lorentz violating coupling parameters by employing polarized or unpolarized sources composed of different fermionic species. Furthermore, positioning the source in close proximity to the hydrogen maser or spin-precession apparatus would enable the investigation of shorter-range spin-dependent interactions mediated by heavier scalar particles. However, at sufficiently small source-detector separations, the Casimir force \cite{Casimir:1948dh}, arising from quantum fluctuations of the electromagnetic field, can become a significant background. Careful characterization and mitigation of Casimir-related effects, along with other short-range systematic backgrounds \cite{Sushkov:2011md,Galanis:2026mxf}, would therefore be essential for achieving the desired sensitivity to these interactions.

Atomic and molecular precision experiments have played a crucial role in advancing fundamental physics, particularly through the development of highly precise atomic clocks and searches for permanent electric dipole moments. In this work, we explore a complementary approach by demonstrating how similar experimental platforms, supplemented with suitably configured external sources, can be employed to probe new physics beyond the SM. In particular, hydrogen maser spectroscopy and atomic and molecular spin-precession measurements provide promising avenues for investigating scalar-mediated spin-dependent interactions and possible violations of Lorentz and CPT symmetries. The ability to control source configurations, interaction distances, and experimental backgrounds offers opportunities to explore previously unconstrained parameter space. Our study thus highlights the broader potential of atomic and molecular precision measurements as versatile probes of ultralight scalar fields and fundamental spacetime symmetries.

\section*{Acknowledgements}
D. C. and T. K. P. are supported by the STFC under Grant No. ST/T001011/1 and No. ST/X003167/1. This article
is based on the work from COST Actions COSMIC
WISPers CA21106 and BridgeQG CA23130, supported
by COST (European Cooperation in Science and
Technology).

\newpage
\appendix
\section{Possible source geometry}\label{appendix1}
Several representative external source object configurations are possible for probing scalar-mediated LV and we discuss them in the following.

For a point-like source i.e., the source is sufficiently small compared with both its distance from the detector and the scalar Compton wavelength, we can write the number density as \cite{Carenza:2025jwn}
\begin{equation}
n_N(\mathbf{r}^\prime)=N_s\delta^3(\mathbf{r}^\prime),
\label{eq:aa5}
\end{equation}
where $N_s$ is the total nucleon number. Therefore, Eq.~\eqref{eq:a4} becomes
\begin{equation}
I_s(r)=\frac{N_s}{4\pi r}e^{-m_\phi r}.
\label{eq:aa6}
\end{equation}
If the source-detector distance is $d$, then Eq.~\eqref{eq:aa6} becomes
\begin{equation}
I_s(d)=\frac{N_s}{4\pi d} e^{-m_\phi d}.  
\label{eq:aa7}
\end{equation}
For a more realistic spherical source, we can consider a sphere of radius $R$, with uniform nucleon density $n_N$ and the total nucleon number $N_s=(4\pi R^3/3) n_N$. Hence, using Eq.~\eqref{eq:a4}, we obtain \cite{Carenza:2025jwn}
\begin{eqnarray}
I_s(r)&=&\frac{n_N}{4\pi}\int {r^\prime}^2 dr^\prime \int d\Omega^\prime \frac{e^{-m_\phi|\mathbf{r}-\mathbf{r}^\prime|}}{|\mathbf{r}-\mathbf{r}^\prime|}\nonumber\\
&=&\frac{N_s}{4\pi r}e^{-m_\phi r}\mathcal{F}(m_\phi R),~~~~~~~\mathcal{F}(m_\phi R)=\frac{3(m_\phi R\cosh (m_\phi R)-\sinh(m_\phi R))}{(m_\phi R)^3},
\label{eq:aa8}
\end{eqnarray}
where in the limit $m_\phi R\to 0$, $\mathcal{F}(m_\phi R)\to 1$, which corresponds to the scalar Compton wavelength larger than the source radius and the source integral reduces to its point source result (cf. Eq.~\eqref{eq:aa7}). If $d$ denotes the gap between the surface of the sphere and the point-like hydrogen detector then Eq.~\eqref{eq:aa8} becomes
\begin{equation}
I_s(d)=\frac{N_s}{4\pi(R+d)}e^{-m_\phi(R+d)}\mathcal{F}(m_\phi R).
\label{eq:aa9}    
\end{equation}

Another relevant geometry for our proposed experiment can be chosen as finite cylindrical plate of radius $R$ and thickness $t_t$. We can put the hydrogen detector on its symmetry axis, a distance $d$ from the nearest face. Using cylindrical coordinates, we can write Eq.~\eqref{eq:a4} as \cite{Croon:2026ake}
\begin{eqnarray}
I_s(d)&=&\frac{n_N}{4\pi}\int^{d+t_t}_d dz\int ^R_0 \rho d\rho \int ^{2\pi} _0 d\varphi \frac{e^{-m_\phi\sqrt{\rho^2+z^2}}}{\sqrt{\rho^2+z^2}}\nonumber\\
&=& \frac{n_N}{2m_\phi}\int ^{d+t_t}_d dz\Big[e^{-m_\phi z}-e^{-m_\phi \sqrt{z^2+R^2}}\Big].
\label{eq:aa10}    
\end{eqnarray}
In the limit where the slab is taken to be infinitely extended, $R\to\infty$, Eq.~\eqref{eq:a10} simplifies to
\begin{equation}
I_s(d)=\frac{n_N}{2m_\phi^2}e^{-m_\phi d}(1-e^{-m_\phi t_t}).   \label{eq:aa11}   
\end{equation}
This configuration corresponds to a point-like hydrogen detector whose dimension is small compared to that of the source slab, so that the field across the detector can be approximated as uniform. From the detector's perspective, the source therefore behaves effectively as an infinite plane.

\section{Near and far switching of the scalar source}\label{appendix2}
Suppose the scalar source is placed at distances $d_n$ and $d_f$ from the hydrogen maser in the near and far configurations, respectively. The corresponding source integrals can then be denoted by $I_n=I_s(d_n)$ and $I_f=I_s(d_f)$ for the two source positions.

Experimentally, we can track $\nu^n_{\rm{Zeeman}}(t)$ and $\nu^f_{\rm{Zeeman}}(t)$ over time and look for any sidereal modulation. Therefore, we can construct 
\begin{equation}
\Delta\nu(t)=\nu^n_{\rm{Zeeman}}(t)-\nu^f_{\rm{Zeeman}}(t),
 \label{eq:aa12}   
\end{equation}
which removes any constant drift. The source integral will change accordingly and Eq.~\eqref{eq:aa12} reduces to
\begin{equation}
2\pi \Delta\nu_\mathrm{Zeeman}(t)=g_N \Delta I_s \Big[\sin\chi\Big(\tilde{\mathcal{J}}_X\cos(\omega_\oplus t)+\tilde{\mathcal{J}}_Y\sin(\omega_\oplus t)\Big)+\tilde{\mathcal{J}}_Z\cos\chi\Big],   
 \label{eq:aa13} 
\end{equation}
where
\begin{equation}
\Delta I_s=I_n-I_f= \frac{n_N}{2m_\phi} \Big[\int ^{d_n+t_t}_{d_n}dz (e^{-m_\phi z}-e^{-m_\phi\sqrt{z^2+R^2}})-\int ^{d_f+t_t}_{d_f}dz(e^{-m_\phi z}-e^{-m_\phi\sqrt{z^2+R^2}})\Big].
\label{eq:aa14} 
\end{equation}
In the large plate limit $R\gg d_n, d_f,t_t,\lambda_\phi$, where $\lambda_\phi=1/m_\phi$ and for a point-like detector, we can write Eq.~\eqref{eq:aa14} 
\begin{equation}
\Delta I_s\simeq \frac{n_N}{2m^2_\phi} (1-e^{-m_\phi t_t})(e^{-m_\phi d_n}-e^{-m_\phi d_f}).   
\label{eq:aa15} 
\end{equation}
If the far position is many Compton wavelengths away, i.e., $m_\phi d_f\gg 1$, then Eq.~\eqref{eq:aa15} becomes
\begin{equation}
\Delta I_s\simeq \frac{n_N}{2m^2_\phi}e^{-m_\phi d_n}(1-e^{-m_\phi t_t}).
\label{eq:aa16}
\end{equation}
Thus, in the large plate and far position limit, we can write Eq.~\eqref{eq:aa13} as
\begin{align}
\mathcal{A}_{nf}(t)&=2\pi \Delta\nu_{\rm{Zeeman}}(t)\nonumber\\
&\simeq \frac{g_N n_N}{2m^2_\phi}e^{-m_\phi d_n}(1-e^{-m_\phi t_t})\Big[\sin\chi\Big(\tilde{\mathcal{J}}_X\cos(\omega_\oplus t)+\tilde{\mathcal{J}}_Y\sin(\omega_\oplus t)\Big)+\tilde{\mathcal{J}}_Z\cos\chi\Big].
\label{eq:aa17}
\end{align}
The contribution $\tilde{\mathcal{J}}_Z$ does not cancel in this near-far switching setup and it converts into source-correlated difference. Thus, with this setup, all the couplings of $\tilde{\mathcal{J}_i}$ can be probed without changing the orientation of the quantization axis.

Therefore, the sidereal amplitude of the $\mathcal{A}_{nf}(t)$ becomes
\begin{equation}
\mathcal{A}^\perp_{nf}=\frac{g_N n_N}{2m^2_\phi}e^{-m_\phi d_n}(1-e^{-m_\phi t_t})\sin\chi\sqrt{\tilde{\mathcal{J}}_X^2+\tilde{\mathcal{J}}_Y^2},
\label{eq:aa18}
\end{equation}
whereas the source-correlated constant amplitude becomes
\begin{equation}
\mathcal{A}^Z_{nf}=\frac{g_N n_N}{2m^2_\phi}e^{-m_\phi d_n}(1-e^{-m_\phi t_t})\tilde{\mathcal{J}_Z}\cos\chi.    
\label{eq:aa19}
\end{equation}

\section{Source integral for the finite size detector}\label{appendix3}

Consider a spherical hydrogen storage bulb of radius $a_b$, centered at $\mathbf{r}_c$. A hydrogen atom inside the bulb is located at $\mathbf{x}=\mathbf{r}_c+\mathbf{r}$, $|\mathbf{r}|\leq a_b$, where $\mathbf{r}$ is measured from the center of the bulb. The scalar source factor at an arbitrary observation point $\mathbf{x}$ is
\begin{equation}
I_s(\mathbf{x})=\frac{1}{4\pi u} \int_{V_s} d^3x^\prime \rho_s(\mathbf{x}^\prime)  \frac{e^{-m_\phi|\mathbf{x}-\mathbf{x}^\prime|}}{|\mathbf{x}-\mathbf{x}^\prime|},
\label{appc1}
\end{equation}
where $u$ is the atomic mass unit, $\mathbf{x}^\prime$ denotes the position inside the external source and $\rho_s(\mathbf{x}^\prime)$ is its mass density. Since, the hydrogen atoms fill a finite storage volume $V_b$, the quantity entering the measured Zeeman shift is the source factor averaged over the bulb
\begin{equation}
\bar{I}_s=\frac{1}{V_b} \int _{|\mathbf{r}|\leq a_b} d^3 r I_s(\mathbf{r}_c+\mathbf{r}), ~~ V_b=\frac{4\pi a^3_b}{3}.
\label{appc2}
\end{equation}
Using Eq.~\eqref{appc1}, we can write
\begin{equation}
 \bar{I}_s=\frac{1}{4\pi u}\int _{V_s}  d^3x^\prime \rho_s(\mathbf{x}^\prime) \frac{1}{V_b} \int _{|\mathbf{r}|\leq a_b} d^3 r \frac{e^{-m_\phi|\mathbf{r}_c+\mathbf{r}-\mathbf{x}^\prime|}}{|\mathbf{r}_c+\mathbf{r}-\mathbf{x}^\prime|}.
 \label{appc3}
\end{equation}
Considering the entire source is outside the bulb i.e., $|\mathbf{x}^\prime-\mathbf{r}_c|>a_b$, we obtain
\begin{equation}
\frac{1}{V_b} \int _{|\mathbf{r}|\leq a_b} d^3 r \frac{e^{-m_\phi|\mathbf{r}_c+\mathbf{r}-\mathbf{x}^\prime|}}{|\mathbf{r}_c+\mathbf{r}-\mathbf{x}^\prime|}=\frac{e^{-m_\phi|\mathbf{r}_c-\mathbf{x}^\prime|}}{|\mathbf{r}_c-\mathbf{x}^\prime|} \mathcal{F}_b(m_\phi a_b),
 \label{appc4}
\end{equation}
where $\mathcal{F}_b(m_\phi a_b)$ is given in Eq.~\eqref{eq:a32}. Therefore, Eq.~\eqref{appc3} becomes
\begin{equation}
\bar{I}_s=  \mathcal{F}(m_\phi a_b) \frac{1}{4\pi u}\int _{V_s}  d^3x^\prime \rho_s(\mathbf{x}^\prime) \frac{e^{-m_\phi|\mathbf{r}_c-\mathbf{x}^\prime|}}{|\mathbf{r}_c-\mathbf{x}^\prime|}=\mathcal{F}_b(m_\phi a_b)I_s(\mathbf{r}_c).   
\end{equation}

\bibliographystyle{JHEP}
\bibliography{reference}
\end{document}